\documentclass[a4paper,10pt,twoside]{revtex4}
\usepackage[utf8]{inputenc}%
\usepackage[left=1in,right=1in,top=0.8in,bottom=0.8in,includefoot,includehead,headheight=13.6pt]{geometry}
\usepackage{amsmath}
\usepackage[makeroom]{cancel}
\usepackage{amssymb}%
\usepackage{graphicx}
\usepackage{comment}
\usepackage{color} 
\usepackage[normalem]{ulem}

\newcommand{\eqn}{\begin{eqnarray}}
\newcommand{\eqnend}{\end{eqnarray}}
\newcommand{\kt}{k_B T}
\newcommand{\bs}[1]{\boldsymbol{#1}}

\newcommand{\pare}[1]{\left( #1 \right) }
\newcommand{\fr}[2]{\frac{#1}{#2}}

\newcommand{\corchete}[1]{\left[ #1 \right]}

\newcommand{\ang}[1]{\langle #1 \rangle}
\newcommand{\tex}[1]{\mbox{\scriptsize{#1}}}

\def\mIt{\bs{\mathcal I}}  
\def\mIc{\bs{\mathbb I}_c}  
\def\mI{\bs{\mathbb I}_0}  
\def\bfL{\bs{L}}  
\def\bfs{{\bf s}}  
\def\volt{\widetilde{\mathbb{V}}}              
\def\vol{\mathbb{V}}              
\def\volm{\widetilde{\mathbb{V}}_m}              
\def\rhom{\bs{\rho}_m}
\def\bS{\bs{S}}
\def\bJ{\bs{J}}
\def\Sm{\bs{\mathcal{S}}}            
\def\Jm{\bs{\mathcal{J}}}

\def\bP{{\bs P}}              %
\def\bu{{\bs u}}              %
\def\bU{{\bs U}}              %
\def\bv{{\bs v}}              %
\def\br{{\bs r}}              %
\def\bF{{\bs F}}              %
\def\inf{{\bs F}^{\prime}}              %
\def\n12{n+\fr{1}{2}}
\def\dt{\Delta t}
\def\S{\bs{S}}

\def\bq{\bs{q}}
\def\bd{\bs{s}}

\def\J{\bs{J}}
\def\bl{\bs{\lambda}}
\def\bL{\bs{\Lambda}}

\graphicspath{{images/}}

\begin{document}
\title{A multiblob approach to colloidal hydrodynamics with inherent lubrication}

\author{Adolfo V\'azquez-Quesada, Florencio Balboa Usabiaga}
\affiliation{Departamento de F\'{\i}sica Teórica de la Materia Condensada, Universidad Autónoma de Madrid, 
  28049 Madrid, Spain.}
\author{Rafael Delgado-Buscalioni}
\affiliation{Departamento de F\'{\i}sica Teórica de la Materia Condensada, Universidad Autónoma de Madrid, 
and Institute for Condensed Matter Physics (IFIMAC). 28049 Madrid, Spain}
\email{rafael.delgado@uam.es}

\begin{abstract}
  This  work   presents  an  intermediate  resolution   model  of  the
  hydrodynamics   of   colloidal    particles   based   on   a   mixed
  Eulerian-Lagrangian formulation.  The particle is constructed with a
  small set of overlapping  Peskin's Immersed Boundary kernels (blobs)
  which  are  held  together  by   springs  to  build  up  a  particle
  impenetrable core. Here, we used  12 blobs placed in the vertexes of
  an  icosahedron with  an  extra  one in  its  center.  Although  the
  particle  surface  is not  explicitly  resolved,  we  show that  the
  short-distance hydrodynamic  responses (flow profiles, translational
  and  rotational mobilities, lubrication,  etc) agree  with spherical
  colloids  and  provide  consistent  effective radii.   A  remarkable
  property  of  the  present  multiblob  model is  that  it  naturally
  presents a ``divergent''  lubrication force at finite inter-particle
  distance.  This  permits to resolve the large  viscosity increase at
  dense colloidal volume  fractions. The intermediate resolution model
  is  able  to recover  highly  non-trivial (many-body)  hydrodynamics
  using {\em small particles} whose radii are similar to the grid size
  $h$ (in the range $[1.6-3.2]\,h$).  Considering that the cost of the
  embedding fluid phase  scales like the cube of  the particle radius,
  this result brings about  a significant computational speed-up.  Our
  code {\em Fluam} works in  Graphics Processor Units (GPU's) and uses
  Fast  Fourier  Transform  for  the  Poisson  solver,  which  further
  improves its efficiency.
\end{abstract}
\maketitle

\section{Introduction}

The hydrodynamic interaction between  colloidal molecules is a central
problem  in   complex  liquid  physics  which   percolates  over  many
disciplines  and  applied research:  from  low Reynolds  hydrodynamics
\cite{KimKarrila,Lowen2012}   to   turbulent   particle   laden   flow
\cite{Eaton2009,Cate2004}.   There is  a solid  theoretical  corpus on
hydrodynamic    interactions    in     the    low    Reynolds    limit
\cite{KimKarrila,Dhont_book1996,Happel-Brenner-book}  which  over  the
years  have  derived most  of  the  closed  analytic forms  for  the
friction and mutual mobilities between pairs of colloidal particles of
different shapes as  a function of their distance  (see e.g. citations
in  Refs.  \cite{Jens2013,Kumar_tesis}).   Colloidal hydrodynamics  is
however a many-body problem  and many body effects become significant
over a  range of colloidal distances \cite{KimKarrila},  which can even
induce      synchronization     of      many      particles     motion
\cite{Cicuta2012}. Computer  simulation is  a key tool  to disentangle
hydrodynamic  effects on colloidal  suspensions and  in some  sense it
faces  the  same sort  of  difficulty  or  limitations encountered  in
analytically  multipole  expansions   and  multiple  reflection  methods
\cite{KimKarrila,BradySierou2001}.   Hydrodynamic interactions between
macromolecules crucially depend on the average separation between them
and  on  the  solute   structure  and  physical  properties  (surface,
permeability,  elasticity,   etc..).   When  dealing   with  soft  and
permeable molecules such  as linear polymers in a  dilute solution, it
is  enough   to  consider   mobility  relations  based   on  pair-wise
interactions  which  just  include   the  lower  order  terms  of  the
multipolar expansion, valid at large interparticle distances (compared
with  the polymer  radius).  The  same  approach is  valid for  dilute
solution  of  rigid  colloids,  however,  when  in  more  concentrated
solution of volume fraction typically above about $15\%$, higher order
terms in their hydrodynamic  interactions start to become relevant. At
dense  concentrations (above  $30\%$ in  volume) the  average distance
between particles is a fraction of the particle radius and lubrication
forces  have  to  be  resolved.   Lubrication is  undoubtedly  the  most
difficult  issue for  any numerical  solver because  it is  a friction
force which diverges at  zero particle distance.  Physically it arises
from the  pressure needed to squeeze  the fluid between  two very near
particles, so in  principle, the numerical solver needs  to be able to
resolve  the fluid  motion  between  the particles  up  to very  small
scales,  well below  the particle  size. The  hard way  to  solve this
dilemma is  to increase the  resolution of the fluid  numerical solver
around the particle  and of the particle surfaces  themselves or using
accurate  and  expensive techniques,  such  as  the spectral  boundary
method  \cite{MuldowneyHigdon1995}.   However,  due to  the  divergent
nature  of   lubrication  forces,  computer  power   soon  limits  the
resolution and the number of colloids and the accessible time window of
the study.

Resolving lubrication forces by explicitly solving the fluid motion 
across  tiny  regions between the colloids  is  a
difficult and costly  route for any numerical scheme.   In fact, it is
even an impossible one if  the fluid-particle coupling is based on the
Stokes  frictional coupling ansatz  \cite{Duenweg2009},  
because  in  this  case only the  lower order terms
of   the  hydrodynamic   multipolar  expansion   (large  interparticle
distances)  will  be  correctly  captured.  In  these  Stokes-coupling
schemes  a clever  strategy to  by-pass  this bottleneck  has been  to
introduce  the theoretical lubrication  force provided  by lubrication
theory,  {\em  ad  hoc}  into  the scheme  for  particle  interactions
\cite{Duenweg2009,Jens2013}.  This ``patched'' lubrication does not 
arise from the motion of the fluid phase, but surely ensures that the friction between
particle  {\em  pairs}  corresponds  to the  theoretical  expectation.
Three body  effects are neglected, although the  role of
many   body  interactions are screened in very dense  suspensions and probably lessens \cite{Kumar_tesis}. 
In fact, the idea  has been successful in dense solutions
of  colloidal  spheres \cite{Kumar_tesis}.  Extending  this  method to  more
complex  molecules is, however,  not straightforward.  Implementation on
non-spherical   geometries   requires   having  theory   and   analytic
expressions  beforehand.   To  further complicate  things,  it  demands some
far-from-trivial   code  adaptations,   such  as   for   instance,  an
(iterative) search of  the nearest points between two  objects and the
determination   of  tangent   planes:  the   reader  is   refereed  to
Ref. \cite{Jens2013}  for an interesting recent work  in that  direction. Another
issue  to consider is  that lubrication  manifests in  quite different
forms  depending on  the relative  motion  of the  objects (normal  or
tangential  relative  motion)  and   also  on  the  type  of  boundary
conditions  at the surface  (bubbles, drops,  slip on  rigid surfaces,
etc).   Finally,  lubrication  also   involves  translation-rotation
coupling which cannot be ignored in many scenarios \cite{Happel-Brenner-book}.

Despite the inherent  difficulties, direct numerical simulations (DNS)
of lubrication, based on the resolution of the underlying fluid motion
around  the colloids, certainly  targets a  more flexible  and general
scheme.  This  route obviously requires  the imposition of  the proper
boundary conditions  at the colloid  surface: the most  common no-slip
surface, and generalizations allowing  for fluid slip in several ways,
depending  on  the  particular  method and  particle  type  considered
\cite{Chew2002,Ferras2013}.  The computational cost of mutual friction and
eventually lubrication  largely depends on the  scheme used.  Particle
based  methods,  such  as   Smooth  Particle  Hydrodynamics  (SPH)  or
Stochastic Rotation  Dynamics (SRD) \cite{Kapral_rev}  are not adapted
to resolve friction at short distances and require large resolution in
terms of  numbers of  solvent particles per  unit volume  to correctly
reproduce  lubrication forces. 
Indeed, preliminary benchmark simulations comparing colloidal lubrication
with SPH with our Eulerian solver with the Immersed  Boundary Method,
show that SPH requires much more  computational resources to
resolve squeezing flow. Again, a possible  solution is  to separately
deal with  lubrication by introducing  ad hoc theoretical corrections
to pairwise friction forces; this 
idea has been recently implemented in splitting integration schemes
for spherical SPH particles \cite{ElleroSplitting2014} and 
also in Stokesian Dynamics, using the so called
Fast Lubrication Method \cite{Kumar_tesis} for spheroids.

Direct  numerical  simulation of  lubrication  using Eulerian  solvers
(where the  fluid motion  is resolved in  a grid using  finite volume,
finite differences  or the Lattice  Boltzmann method) need  to resolve
the particle  surface where to impose the  desired boundary condition,
which in  practice involves  modifying the flow  over the  nearby fluid
sites. There are several possible strategies. Lattice Boltzmann solvers
have the  possibility of defining collision rules  for the border-sites
between  the   solid  and  the  fluid  which   modify  the  underlying
``microscopic''  LB  distribution   function  in  the  advection  step
\cite{Duenweg2009}  in  ways  that  might represent  no-slip  or  slip
surfaces.  Works  using this method use particle  diameters between 10
and 40 fluid  cells (or lattices) to patch  the lubrication correction
at  colloidal   separations  below   $5\%$  of  the   particle  radius
\cite{Nguyen2002}.  Similar  diameters are reported  in previous works
using    finite    elements    \cite{KimKarrila},    Direct    Forcing
\cite{Uhlmann2005,Breugem,Feng2005},   the   Smooth   Profile   method
\cite{Molina-Yamamoto2013}

Immersed  Boundary  method  \cite{Peskin2002} is  another  referential
method  which has  many variants  such  as the  Direct Forcing  method
\cite{Uhlmann2005,Feng2005} or the Stochastic Immersed Boundary method
\cite{Atzberger2007}.   Direct Forcing,  uses marker  points  over the
particle {\em surface} where to impose the no-slip condition.  Surface
or  boundary conditions  are converted  into volume  forces  which are
spread over  the fluid  cells inside the  local kernel to  ensure zero
relative fluid-particle velocity  in it.  In this work  we carry out a
similar strategy  with an important difference:  our resolution target
is  not the particle  surface. From  previous works  we know  that the
``physical'' volume of  each blob and its hydrodynamic  size (which is
about   one   mesh   size   $h$)   is   determined   by   its   kernel
\cite{Balboa2014,Balboa_tesis}.   The idea  of  a multi-blob  particle
formed  by  overlapping  blob-kernels,  is  then not  to  resolve  any
surface, but to  let the surface (or more  properly, the particle size
and shape) emerge from the  reconstruction of the particle body.  This
subtlety is important  because our goal is to  develop an intermediate
level of particle  resolution, where colloids are of  the same size of
the mesh (the  smallest radius hereby considered is  $R \simeq 1.5 h$)
but  still present  an impenetrable  core and  significant lubrication
forces.  Lubrication forces arise  ``naturally'' from the fluid solver
and quite  importantly,  {\em increase without bounds at finite  inter-particle
  distances}.   This  ``divergence''  of  the  model's  lubrication comes out
from  a nice  property of  single blob  kernels, whose  mutual friction
diverge  at  full  overlap \cite{Balboa2012a,Balboa2014}. This result
opens the  possibility of studying lubrication effects using
rather small particles whose radius is about 1.5$h$. It has to be stressed
that for  fixed accuracy in  the mutual mobility resolution,  a linear
reduction in the particle size  permits a cubic reduction in the fluid
volume, which is certainly  substantial and permits longer simulations
with more particles. On a broader perspective, it also offers
another route to lubrication of complex shapes.

In  this  work the  blob  forming  the  multiblob particles  are  held
together  using  hard harmonic  springs.   This  is  certainly not  an
efficient method if  one is interested in rigid  body {\em motion} and
alternatives         have         been        already         proposed
\cite{Feng2005,Molina-Yamamoto2013,Bhalla2013b}   and  implemented  in
schemes  with  linear accuracy  in  the  time  stepping. However,  the
stresslets  arising from  the  rigid body  {\em  constraint} are  more
difficult to  implement and probably require iterative schemes;  in the present
approach the contribution from the  particle stress is provided by the
connecting internal  springs. The choice of the  model connectivity is
in part  practical, as  one of our  next research goals  is elasticity
effects in  colloid hydrodynamics which despite its  relevance in many
disciplines (notably in biology,  microgels and even in nano-clusters)
has been seldom  considered in the literature (see  the recent paper
by Felderhof \cite{Felderhof2014}).  We recently developed single blob
models  with  arbitrary  compressibility \cite{Balboa_sound}  and  the
elastic  multiblob  naturally generalize  to  particles with  arbitrary
Young modulus.

We  will first  present in  Sec. \ref{sec:model}  the  multiblob model
which in this work has the icosahedron shape. This platonic solid hugs
its inscribed sphere  the most tightly and its  surface area to volume
ratio is the closest to a sphere of the same volume. Using the smallest
icosahedron tessellation, with $12$  vertexes one gets a volume filling
factor (ratio  between the icosahedron  and sphere volume)  of $\simeq
0.61$.   In Section \ref{sec:eq}  discuss the  hydrodynamic properties
and induced stress of one  multiblob from the analysis its equation of
motion.  Then in  Sec.  \ref{sec:mass} we use this  analysis to derive
the  particle volume  and mass,  and momentum  of inertia.   The model
hydrodynamic behavior starts with a study of its Stokeslet and rotlet
response   at  long   distances   (Sec.   \ref{sec:stokes}),   proving
consistency  with Einstein relation  (between the  hydrodynamic radius
and self-diffusion).   We then focus on the  particle hydrodynamics at
closer  distances,  and  study  the  flow velocity  profile  past  the
multiblob,   Sec.   \ref{sec:vel}.   Pair   hydrodynamic  interactions
(normal  and tangential mobilities,  translation-rotation coupling
and  lubrication  curves)  are  studied in  Sec.  \ref{sec:fric}.   To
conclude the tests in Sec.  \ref{sec:stresslet} we study the effective
viscosity  of  an ensemble  of  multiblobs  and  observe an  excellent
agreement with  Batchelor's celebrate result for spheres.   We also show
that the model  can reproduce the viscosity increase  at larger volume
fractions.  We conclude with comments on the model consistency and the
sort of methodological research  this study suggest taking for further
improvements.

\begin{figure}
  \begin{center}
    \includegraphics[scale=0.5]{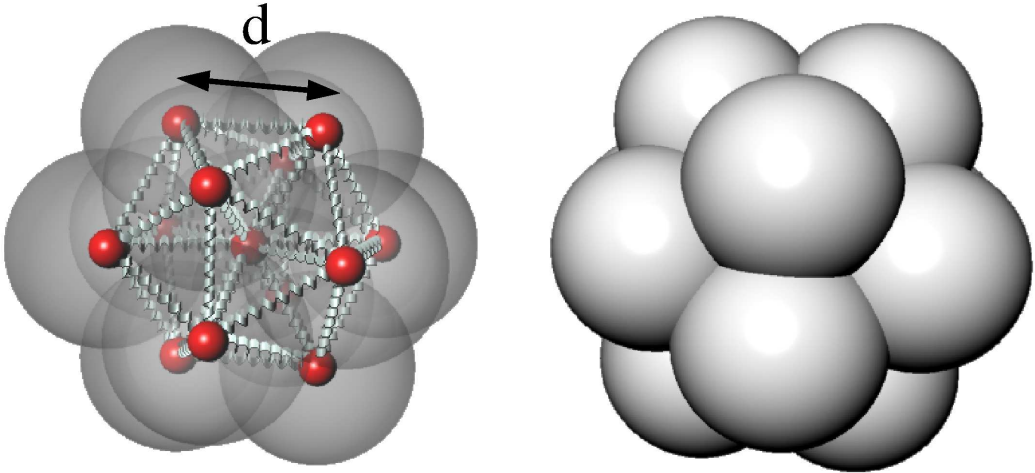}
  \end{center}
  \caption{A multiblob particle formed by $N=13$ blobs placed in
    the 12 vertex of an icosahedron and its center. The size of the particle
    is determined by the distance between vertex, $d$. Blobs centers (red balls in the figure) are held together 
    by hard springs, as indicated in the figure. 
    \label{fig:ico}
  }
\end{figure}

\section{Multiblob particle model}
\label{sec:model}

The multiblob particle consists on a set of $N$ linked blobs to form a
bigger particle. In  this work the blobs are  linked with hard springs
to conform an icosahedron, as shown in Fig. \ref{fig:ico}.  In order to
investigate the effect of possible  leakage of fluid into the particle
domain on  its hydrodynamic properties  (mutual friction, lubrication)
we have  considered two models: an icosahedron with $N=12$ where 
blobs are placed at its vertexes and it is empty inside, and another model
with an extra  blob in the center to form a filled icosahedron with $N=13$ blobs.
The baricenter of the particle $\bq_0$ is given by the 
positions of the blobs $\bq_i$,
\begin{eqnarray}
  \bq_0 &=&\frac{1}{N} \sum_i \bq_i\\
  \bq_i &=& \bq_0 + \bd_i
\end{eqnarray}
The translation velocity of the particle is 
\begin{equation}
  \bU_0 = \frac{1}{N} \sum_{i=1}^{N} \bu_i.
\end{equation}
The particle motion has also degrees of freedom related to
rotation and vibration of blobs around the baricenter. For any blob,
the equilibrium distance $s_i^{eq}$ 
to the icosahedron center is fixed by the spring.
For the blobs in the shell, $s_i^{eq} = a$
where  $a$ is the icosahedron radius 
($a \simeq 0.950\,d$ in terms of inter-vertex distance $d$).
In this work the springs are hard enough to neglect vibration or fluctuations
around the equilibrium sites, i.e. $s_i \simeq s_i^{eq}$.
In this rigid body limit we can write,
\eqn
  \bu_i = \bU_0 + {\bs \Omega} \times \bd_i,
\eqnend

We apply the {\em no-slip} condition to each blob, so the fluid velocity $\bv(\br)$
in the blob domains follows the particle. 
In the continuum formulation the {\em no-slip} condition corresponds to \cite{Balboa2012a,Balboa2014}
\begin{equation}
  \label{ju}
  \bu_i =  \J_i \bv = \int \delta_h(\br-\bq_i) \bv(\br) d\br
\end{equation}
where $\J_i$ is the interpolator operator of blob-$i$
and $\delta_h(\br-\bq_i)$ is the blob kernel, here constructed as a tensorial product
of the 3-point Peskin's function $\Phi(x)$ \cite{Roma1999,Peskin2002,Balboa2012a},
\begin{equation}
  \label{delta}
  \delta_h(\br) = h^{-3} \prod_{\alpha=1}^3 \Phi\corchete{\fr{r_{\alpha}}{h}} 
\end{equation}
with $h$ the regular mesh size used in the present method.
In the discrete  setting, the integral in Eq.  (\ref{ju}) is converted
into a  sum over all  the lattice cells.  The  interpolator $\delta_h$
inherits all  the grid-independence  features of the  3-point Peskin's
function $\Phi(x)$.   In particular: the zeroth and  first moments and
the discrete  $L^2$-norm of the  kernel are grid independent.  The two
first conditions  are $\int \delta_h(\br-\bq)  \br^{n}_{\alpha} dr^3 =
q^{n}_{\alpha}$ for $n=0$ and $n=1$,  while the last  condition means
that  $\int  \delta_h(\br)^2  dr^3  =  \vol^{-1}$, where  for  a  blob
particle   model   in   a  compressible   fluid   \cite{Balboa2012a,Balboa_sound}, $\vol=8\,h^3$ is the volume of fluid occupied by the particle.

We introduce the multiblob interpolator  $\Jm$,
\begin{equation}
  \label{jm}
  \Jm \equiv \frac{1}{N} \sum_{i} \J_i
\end{equation}
which provides the translation multiblob velocity from the underlying fluid velocity field,
\begin{equation}
  \bU_0 = \Jm \bv.
\end{equation}
The adjoint of the blob interpolator $\bJ_i$ is the 
spreading operator which distributes
the force ${\bf  F}_i$ acting  on  each blob 
as a force density to the fluid phase,
${\bf  f}_i(\br)  =\S_i {\bf  F}_i=\delta_h(\br-\bq_i) {\bf F}_i$  (see  \cite{Balboa2014,Balboa_tesis}). 
As before one can introduce the
overall spreading operator
\begin{equation}
  \Sm \equiv \frac{1}{N} \sum_{i} \S_i = \frac{1}{N} \sum_i \delta_h(\br-\bq_i)
\end{equation}
which transmits {\em net} forces on the multiblob particle to the surrounding
fluid.

As stated,  the average and  spreading operators of the  isolated blob
are adjoint and ensure certain important properties \cite{Balboa2014}:
Grid  independence of  the zeroth  and first  moments  $\J_i[1]=1$ and
$\J_i\left[\br-\bq_i\right]=0$  and   constant  norm,  which   can  be
generally expressed as $\J_i  \bP \S_i = \bs{\mathcal{V}}^{-1}$, where
in the compressible setting $\bP^{\alpha,\beta}=\delta_{\alpha,\beta}$
is  just the  identity operator  while, in  the  incompressible limit,
$\bP$ is  the solenoidal  projector which applied  to a  vector field,
provides its  divergence-free component (see  Ref. \cite{Balboa2014}).
For  $\bP=\bs{1}$ (compressible  case)  the inverse  volume matrix  is
independent on the grid and $\bs{\mathcal{V}}^{-1}=\vol^{-1} {\bf 1}$.
In this  incompressible setting, due to  the instantaneous elimination
of  the  sound  modes  \cite{Balboa_tesis,Balboa2014},  the  effective
particle volume  is slightly larger,  $\volt=(3/2) \vol$, consistently
with the  added mass  effect of a  real particle in  an incompressible
fluid  \cite{Zwanzig1975,LandauFL}.   In  the incompressible  discrete
formulation    \cite{Balboa_tesis,Balboa2014}    we    observe    that
$\bs{\mathcal{V}}^{-1}  \approx \volt^{-1} {\bf  1}$ (about $1\%$
deviation). Following  with the blob-multiblob equivalence,  due to the
linearity of $\Jm$  and $\Sm$, it is easy to see  that $\Jm$ and $\Sm$
are also adjoint  and have grid independent zeroth  and first moments,
$\Jm[1]=1$ and  $\Jm \left[\br-\bq_0\right]=0$ (where  $\bq_0$ is the geometric center of the multiblob).  As an extrapolation
of   the  equivalences  with   the  single   blob  case   analyzed  in
Ref. \cite{Balboa_tesis,Balboa2014}, the operator,
\begin{equation}
  \Jm \bP \Sm  = \frac{\sum_{i,j} \J_i \bP \S_j}{N^2} 
\end{equation}
should ideally be proportional to the inverse of the multiblob volume, $\volm$.
Moreover, we show in subsequent sections that, for our isotropic body
this operator present small deviations from a scalar, i.e.,
$\Jm\bP\Sm \simeq (\volm)^{-1} {\bs 1}$.

\section{Multiblob dynamics}
\label{sec:eq}
The particle we  consider in this work is not completely rigid. It is
formed by blobs  connected by hard springs so  formally it corresponds
to the  type of flexible structures  treated in immersed boundary  
methods \cite{Peskin2002}. The interblob springs are however 
stiff enough to ensure that the amplitude of particle vibrations are small
which in practice, it means that the frequency of the springs 
is faster than the rate of viscous fluid momentum distribution around the particle.
Using springs to connect the blobs and form multiblobs permits us to use
the same solvers we have recently developed for single blob dynamics in different hydrodynamic regimes \cite{Balboa_tesis}.
For this work we have mainly used the Inertial Coupling method for particles in incompressible fluid \cite{Balboa2014} (which
includes both fluid and particle inertia) and also the Fluctuating Immersed Boundary method \cite{Stokeslimit},
adapted to the Stokesian limit, where the particle and fluid inertia is absent.
In this section we discuss the hydrodynamic response and properties of one multiblob, arising from 
their equations of motion.

\subsection{Translational motion}
The velocity of one blob $i$ in the particle can
can be decomposed in the particle translation velocity $\bU_0$ 
and a peculiar velocity $\bu^{\prime}_i$, 
\begin{equation}
  \bu_i =\bU_0 + \bu^{\prime}_i,
\end{equation}
Neglecting the small vibrations of the blobs, the  instantaneous angular velocity ${\bs \Omega}$ of the body
can be determined from $\bu^{\prime}_i= {\bs \Omega} \times \bd_i$, which is an
exact relation for a rigid body.

In our model the incompressible fluid velocity  field with 
a single multiblob particle evolves according to,
\begin{equation}
  \label{vf}
  \rho_f \frac{\partial \bv}{\partial t} =-\nabla \pi -\nabla \cdot \bs{\sigma}_{f} - \sum_{j} \S_{j} \bl_{j} = \bP\left(\nabla\cdot {\bs \sigma}_{f}
  - \sum_{j} \S_{j} \bl_{j} \right)
\end{equation}
where $\pi$ is the  hydrodynamic pressure or the Lagrangian multiplier
ensuring  the incompressibility condition  $\nabla \cdot  {\bv}=0$ and
$\bs{\sigma}_{f}=  -\eta \bs{E} +\tilde{\bs{\sigma}} +\rho_f \bv  \bv$ 
contains the viscous stress constructed with the symmetric part of the fluid strain
($\bs{E}\equiv (\nabla \bv +\nabla \bv^{T})/2$) and
the  convective inertia, given by the dyadic  $\rho_f \bv \bv$.
Hydrodynamic fluctuations (either in compressible  or incompressible flow) can be
also  included via a fluctuating stress  $\tilde{\bs \sigma}$  as
explained  in  Refs.  \cite{Balboa2012a,Balboa2014},  whose  validity 
is restricted to small departures  from the equilibrium state (Gaussian
statistics).

The LHS of the momentum  density equation can be also formally written
without the hydrodynamic pressure $\pi$ using the operator $\bP$ which
projects any field on its  solenoidal sub-space $\nabla \cdot \bP {\bf
  v}=0$. 

The force $\bl_{i}$ is a Lagrangian multiplier ensuring the {\em no-slip} condition on each blob,
\begin{equation}
  \bu_{i}= \J(\bq_i) \bv = \J_i \bv .
\end{equation}
The fluid-particle force $\bl_i$ is also present in the blob equation of motion (see Ref. \cite{Balboa_tesis}),
\begin{equation}
  \label{blob1}
  \bl_i=  m_{e,i} \dot \bu_i - {\bf f}_i \,,
\end{equation}
where $m_{e,i}$ is its excess mass and the total mass of the {\em isolated} blob is
\begin{equation}
  \label{mblob1}
  m_i=\rho_f\tilde{\vol} + m_{e,i}.
\end{equation}
The blob equation of motion can be also obtained by noting that
\begin{equation}
  \frac{d\bu_i}{dt}= \J_i \left (\frac{\partial \bv}{\partial t}\right)+ {\bf a}_i
\end{equation}
where ${\bf a}_i\equiv \bu_i \cdot \nabla_{\bq} \J_i \bv$ is the convective acceleration arising from the material derivative relative to the blob translation (see \cite{Balboa2014,Balboa_sound}).

Applying $\J_{i}$ in Eq. (\ref{vf}) leads to,
\begin{equation}
  \label{ui}
  \rho_f \dot \bu_i = -\sum_{j} \J_{i} \bP \S_{j} \bl_{j} -\J_i\bP \nabla\cdot {\bs \sigma}_{f}  +\rho_f {\bf a}_i
\end{equation}
Introducing Eq. (\ref{blob1}) and the density matrix operator,
\begin{equation}
  \label{rhoi}
        {\bs \rho}_{i,j} \equiv \rho_f \delta_{ij} +  \J_i\bP \S_j m_{e,j} \nonumber
\end{equation}
Eq. (\ref{ui}) can be expressed explicitly in terms of the set of blob accelerations,
\begin{equation}
  \label{ui2}
  \sum_j {\bs \rho}_{i,j} \dot \bu_j =  
  \sum_{j} \J_{i} \bP \S_{j} {\bf f}_{j} -\J_i\bP \nabla\cdot {\bs \sigma}_{f}  +\rho_f {\bf a}_i \,.
\end{equation}

The forces acting on each blob can be decomposed in a net contribution to 
the particle translation and peculiar components, i.e.
\begin{eqnarray}
  \bl_i &=& \frac{1}{N} \bL_0 +\bl_i^{\prime},\\
  \bL_0  &=& \sum_{i} m_{e,i} \dot \bu_i - \bF_0,\\
     {\bf f}_i &=& \frac{1}{N} \bF_0 +{\bf f}^{\prime}_i.
\end{eqnarray}
The total non-hydrodynamic force acting on the particle is $\bF_0$ and 
the peculiar force $\inf_i$ sums to zero ($\sum_i {\bf f}^{\prime}_i=0$)
and contains contributions from external torque and internal force (springs).
The fluid contributes with a net $\bL_0$ force on the multiblob, which
equals $-\bF_0$ if the particle has no inertia ($m_{e,i}=0$ for each blob).

The equation of motion of the multiblob geometric center $\bq_0 = (1/N) \sum_i \bq_i$
is obtained by summing over the $N$ blobs in Eq. (\ref{ui}). If all the excess masses are similar $m_{e,i}=m_e$ 
the motion of the geometric and the center of mass coincide
and its acceleration $\dot \bU_0$ is given by,
\begin{equation}
  \label{u0}
  \rhom \dot\bU_0 =  \Jm\bP\Sm \,\bF_0
  - \Jm \bP \left[\nabla\cdot {\bs \sigma}_{f}
    + \sum_j \S_j \bl_j^{\prime}\right]
  + \rho_f {\bf a}_0.
\end{equation}
where we have defined the average multiblob density as,
\begin{equation}
  \label{rhom}
  \rhom = \rho_f {\bs 1} + N \Jm\bP \Sm m_e = \frac{1}{N}\sum_{i,j} {\bs \rho}_{i,j}
\end{equation}

Equation (\ref{u0}) reflects that the particle translation velocity is
not only  affected by the  net force but  also by the  distribution of
peculiar   blob   forces.   Physically   this   is   related  to   the
translation-rotation  coupling  \cite{Happel-Brenner-book}  or in  the
case    of    elastic    particles   translation-vibration    coupling
\cite{Felderhof2014},  much less  studied  in the  literature.  In  an
isotropic body  $\Jm\bP\Sm$ should be  a scalar \cite{Happel-Brenner-book}, 
however deviations might occur in the discrete setup
leading to spurious forces. We will come back  to this issue later.

\subsection{Rotational motion}
\label{subsec:rotationalMotion}
The equation of motion for the multiblob angular velocity ${\bs \Omega}$ can be derived 
from Eq. (\ref{ui2}), by performing the vector product $\bfs_i\times$ summing over the blobs of the multiblob.
In the most general case (a multiblob with arbitrary excess mass distribution $m_{e,i}$),
\begin{equation}
  \label{rot1}
  \sum_{i,j} \bfs_i \times {\bs \rho}_{i,j} \dot \bu_i =  \sum_{j} \bfs_i \times \J_{i} \bP \S_{j} {\bf f}_{j} -  \sum_i \bfs_i \times \J_i \bP \nabla\cdot {\bs \sigma}_{f}  +\rho_f \sum_i \bfs_i \times {\bf a}_i
\end{equation}
This equation contains a significant amount of well known hydro-mechanical couplings:
such as crossed inertial terms which might induce rotation from translational accelerations,
torques arising from non-homogeneous mass distributions and 
also in the Stokes limit, crossed rotational-translational  mobilities.
In the present work we shall
just focus on the inertia-less, neutrally buoyant case ($m_{e,i}=0$ and $\bs{\rho}_{i,j}=\rho_f\delta_{i,j}$) 
and will consider isotropic bodies,
where the translation-rotational mobility coupling is absent. Consider
a pure rotational motion of the icosahedron induced by an external torque resulting from a set of forces 
on the blobs given by ${\bf f}_i = {\bs \alpha}\times \bfs_i$, where ${\bs \alpha}$ is a constant vector.
The resulting torque is,
\begin{equation}
  \label{torque}
        {\bs \tau} = \sum_{i} \bfs_i \times {\bf f}_i = \sum_{i} \bfs_i \times {\bs \alpha}\times \bfs_i = \mI {\bs \alpha}
\end{equation}
where we have defined the tensor,
\begin{equation}
  \label{miner0}
  \mI=\sum_i \left[ s_i^2 {\bs 1} - \bfs_i \bfs_i \right]
\end{equation}
Applying the time derivative to $\bu_i=\bU_0 +{\bs \Omega}\times \bfs_i$ and using that ${\bf f}_i={\bs \alpha}\times \bfs_i$,
one gets, after some manipulations in Eq. (\ref{rot1}),
\begin{equation}
  \label{rot1b}
  \rho_f \left[ \mI \dot {\bs \Omega} + \sum_{i,j} \bfs_i \times \left({\bs \Omega} \times \left({\bs \Omega} \times \bfs_j\right)\right)  
\right]=   \mIc \mI^{-1} {\bs \tau}  - \sum_i \bfs_i \times \J_i \bP \nabla\cdot {\bs \sigma}_{f}  + \rho_f \sum_i \bfs_i \times {\bf a}_i,
\end{equation}
The torque arising from the convective force $\sum_i \bfs_i \times {\bf a}_i$ can only be significant at large particle
Reynolds number, which shall not be considered in this work. The term $\sum_{i,j} \rho_f  \bfs_i (\times {\bs \Omega} (\times {\bs \Omega} \times \bfs_j))$ vanishes for isotropic homogeneous bodies and shall not be considered hereafter. 
To  obtain the first term on the RHS of Eq. \ref{rot1b}, $\mIc \mI^{-1} {\bs \tau}$
we have write it as $\sum_{i,j} \bfs_i \times (\bJ_i\bP\bS_j (\bs{\alpha} \times \bfs_j))=\mIc \bs{\alpha}$ (see below) and then introduced the relation for the external torque  ${\bs \alpha} = \mI^{-1} {\bs \tau}$. Using the above assumptions, Equation (\ref{rot1}) becomes,
\begin{equation}
  \label{rotnb}
  \rho_f\, \mI \dot {\bs \Omega}  = \mIc \mI^{-1} {\bs \tau}  - \sum_i \bfs_i \times \J_i \bP \nabla\cdot {\bs \sigma}_{f}.
\end{equation}
The  second order  tensor  $\mIc$ can  be  expressed in  terms of  the
Levi-Civita  third order tensor  $\epsilon^{\alpha,\beta,\gamma}$ used
in cross-product manipulations (note  that Einstein convention is used
for  superindexes  which  corresponds  to  spatial  directions,  while
subindexes to the blobs)
\begin{equation}
\label{mict}
\mIc^{\alpha,\mu} \equiv \sum_{i,j} \epsilon^{\alpha,\beta,\gamma} \epsilon^{\lambda,\mu,\nu} s_i^{\beta} s_j^{\nu} (J_i \bP S_j)^{\gamma,\lambda}
\end{equation}
which can be further re-expressed using operations with second order tensors
\cite{note}.  In  the incompressible fluid case $\mIc$  is difficult to
calculate  because the tensorial character  of $\bJ_i\bP  \bS_j$
modifies the direction of any vector which is applied to.
The compressible case ($\bP={\bf 1}$) is simpler because $\bJ_i \bS_j$ is an scalar so
$\bfs_i \times (\bJ_i\bS_j (\bs{\alpha} \times \bfs_j))= \bJ_i\bS_j (\bfs_i \times (\bs{\alpha} \times \bfs_j))$ and,
\begin{eqnarray}
  \label{minerc}
  \mIc&=&\sum_{i,j} \bJ_i \bS_j \, \left[ \bfs_i \cdot \bfs_j {\bs 1} - \bfs_j \bfs_i \right].\\
\end{eqnarray}

We finally note that $\dot {\bfL}= \rho_f \mI \dot {\bs \Omega}$ is just the time derivative of the 
angular momentum density of the (neutrally buoyant) particle 
\begin{equation}
  \label{angularmom}
  \bfL = \rho_f \sum_i \bfs_i \times \bu_i.
\end{equation}
Equation (\ref{rot1b}) or (\ref{rotnb}) permits to evaluate the multiblob momentum of inertia, 
which will be presented in Sec. \ref{sec:mass} along with
an alternative derivation based on the equipartition theorem.

\subsection{Stresslet}
\label{stresslet}
Unlike the single blob particle model, the multiblob is able
to create stress on the fluid due to its cohesive forces. This can clearly observed by expressing 
the particle force density contribution to the fluid momentum
equation \ref{vf} in the form of the divergence of a particle pressure tensor.
To that end we expand the spreading function of a single blob in a Taylor
series around $\br-\bq_0$,
\begin{equation}
  \S\left(\br-\bq_i\right)= \S\left(\br-\bq_0-{\bf s}_i\right) =
  \S\left(\br-\bq_0\right) -\nabla_{\br} \S(\br-\bq_0) \cdot {\bf s}_i +\mathrm{h.o.t.}
\end{equation}
So the particle contribution in Eq. \ref{vf} becomes,
\begin{equation}
  \label{sexp}
  -\sum_i \S(\br-\bq_i) {\bl}_{i} = 
  -\S\left(\br-\bq_0\right) \bL_0 + \nabla_{\br} \cdot \left[ 
    \S(\br-\bq_0) \sum_{i=1}^N {\bf s}_i\, \bl^{\prime}_i\right]  +\mathrm{h.o.t.}
\end{equation}
The first term in Eq. (\ref{sexp}) is the monopole contribution of the
multiblob  particle (net  force) while  the dipolar  contribution only
involves peculiar  (fluid-particle) forces $\bl^{\prime}$  (as $\sum_i
{\bf s}_i =0$). Higher order  terms in the expansion (\ref{sexp}) become
relevant  if  the  particle  size  is  comparable  with  the  smallest
wavelength $2\pi/k$ of  the fluid velocity spectra.  This  can be seen
by performing a  Fourier transform (in $\br$) over  the full expansion
of Eq.   \ref{sexp} and realizing that  the next term  in the pressure
tensor is proportional to $(i/2)  {\bf k} {\bf s}$. In most colloidal
applications,  the characteristic length  of the  flow is  much larger
than the colloidal size, $ka<<1$ (we were not able to find a study on the
effect  of  higher order  multipoles  in  the  viscosity of  colloidal
fluids).  The dipole term in Eq. (\ref{sexp}), whose dyadic is usually
decomposed in a rotational (skew-symmetric part) and a stresslet (zero
trace,   symmetric  part)   \cite{KimKarrila},  takes   now   the  form
$-\nabla\cdot  \bs{\sigma}_{p}$  where  the components  of  particle
pressure tensor are,
\begin{equation}
  \label{presp}
  \sigma_{p}^{\alpha,\beta} = -\S(\br-\bq_0) \sum_{i=1}^N \langle s_i^{\alpha} {\bf f}^{\prime\,p\beta}_{i}\rangle
\end{equation}
Were we have assumed a case where inertia is negligible so that $\bl_i
=  -{\bf  f}_i$  (see  Eq.   \ref{blob1}). The average is made over the
fast dynamics of the multiblob, which depending on the application (or relevant time scale) 
might  include the vibrational modes only, or the translational and rotational degrees of freedom as well. 
The  contribution  of  one particle  to the  pressure tensor  is thus  sustained by  its internal
virial term.  

Equation  (\ref{presp}) shows  that  the dipolar  contribution of  the
multiblob model is  the sum of $N$ dipoles ${\bf  s}_i {\bf f}_i$.  In
the present  model the position  of the blob  $i$ with respect  of the
body center $\bq_0$ can fluctuate  ${\bf s}_i = {\bf a}_i +\delta {\bf
  s}_i$,  (with  ${\bf  a}_i =  a  {\bf  n_i}$  the radius  vector  to
embedding  sphere). Thus,  the particle  virial contain  a  rigid body
contribution ${\bf  a}_i {\bf  f}_i$ and a  elastic one $  \delta {\bf
  s}_i {\bf  f}_i$.  The relevance  of fluctuations of the  body shape is
determined by  the ratio $\delta s_i/a$.  Here,  the displacement from
the equilibrium position is $\delta  s_i \sim f_i/k_{sp}$.  In a fluid
at  rest,  the  equipartition  of  the spring  energy  in  equilibrium
establishes  that $\delta  s_i \sim  \left(\kt/k_{sp}\right)^{1/2}$ and
the ratio $\delta s_i/a \sim O(\kt/k_{sp} a^2)^{1/2}$.  Under straining
motion $f_i \simeq 6\pi \eta E  a^2$ (where $E$ is the shear rate) and
in terms of  the Peclet number $\mbox{Pe} = 6\pi\eta  E a^3/\kt$ the relevance
of the  shape fluctuations scales like $\delta  s_i/a \sim \mathrm{Pe}
(\kt/k_{sp}a^2)$.  In  the present simulations we have  used very stiff
springs  which  ensure  these  non-dimensional groups  are  negligible
small.

The stress released by the particle is directly related to the local stress of the fluid.
For  an  inertia-less and  force-free  particle  in straining  fluid
motion, the  force on the blob $i$ is the local surface  traction of the fluid 
${\bf f}_i^{\prime}  =  \left(\pi{\bs 1}+{\bs  \sigma}_{f}\right)  \cdot  {\bf n}_i  dA_i$  where  $dA_i$ is  the
surface element  of the blob  and ${\bf n}_i  = {\bf s}_i/s_i$  is the
surface vector.  In the continuum limit of the body space coordinates,
such  relation  would provide  the  proper  continuum  version of  the
particle stress, ${\bs \sigma}_{p}= \oint {\bf s} (\pi {\bs 1}+ {\bs
  \sigma}_{f})  \cdot {\bf  n}  dA$ \cite{KimKarrila}.   It is possible to map our discrete formulation to
the continuum  formulae, however in this  work we will analyze the
consistency  of  the  multiblob  stresslet  by  considering  its  most
prominent effect in  the flow, which is the  increase of the effective
viscosity of a multiblob colloid suspension (see Sec. \ref{stresslet}).

\section{Multiblob volume, mass and moment of inertia}
\label{sec:mass}

In this section we use the relations obtained in Sec. \ref{sec:eq}
to derive several properties of the multiblob, such as its mass and moment of inertia.
By ascribing the moment of inertia of the multiblob to that of a sphere
(which in fact coincides with that of a solid icosahedron) we will obtain a first 
measure of the multiblob radius.
To derive the mass and moment of inertia of the multiblob particle we analyze
its inertia, from the transient of Eq. (\ref{u0}) and (\ref{rotnb}). Then, 
we show that the {\em inertial} evaluations are consistent 
with the {\em thermal} ones based on the equipartition of the translational and rotational kinetic energy. 

\subsection{Multiblob mass and volume}

\subsubsection{From the particle inertia}
The inertial mass in Eq. (\ref{u0}) can be obtained by considering 
a particle pulled with a constant force $\bF_0$ which is equally distributed over
the forming blobs ${\bf f}_i=\bF_0/N$. If the fluid is ideal 
$\bP \nabla\cdot \bs{\sigma}_{f}=0$ and initially at rest ${\bf a_0} =0$ in Eq. \ref{u0},
the particle mass could be directly obtained from the ratio $M=F_0/\dot U_0$. In practice one can 
reproduce this situation by considering very short times $t\rightarrow 0$ after the force imposition,
i.e., when the fluid still behaves like ideal, well before the onset of the frictional regime.
In the Eulerian (inviscid) limit, the force $\Jm\bP\Sm \bF_0$ depends on the body shape, which can induce lift forces 
(normal to $\bF_0$). However the lift is zero for (non-rotating) isotropic bodies. According to Eq.  \ref{u0}, the mass of 
the multiblob is well defined if the $\Jm\bP\Sm$ operator is proportional to the inverse of the particle volume so
for our isotropic body we expect,
\begin{equation}
  \Jm \bP \Sm \bF_0 = \volm^{-1} \bF_0
\end{equation}
where $\volm$ is the multiblob volume. 
The situation is then similar to the single blob case \cite{Balboa2014}, for
which the relation $\J_i\bP \S_i = \volt^{-1} {\bf 1}$ is exact in the
continuum periodic  setting.  In the discrete  formulation it presents
deviations   of   about   $1\%$    over   the   grid   positions   and
directions \cite{Balboa_tesis}.  In  the multiblob  case we expect  that the
extensive overlap does not substantially alter this quasi-independence
of the particle volume with  the grid.  In principle, the blob kernels
used   to  build  the   multiblob  are   not  constructed   to  ensure
such grid-independence. This kind of optimization still being an
open problem  \cite{Pinelli2010a}.
Albeit,  we have found  that a
set of  overlapping Peskin's 3pt-kernels leads to  small variations in
mass,  volume and hydrodynamic  size of  the multiblob  icosahedron, with
relative deviations  which can be even  smaller than that  of a single
blob (see  Figure \ref{fig:mass}b).  The (isotropic) particle density operator
in Eq. (\ref{rhom}) is then also almost regular, $\rhom \approx \rho_m {\bs 1}$, 
with
$$
\rho_m \approx \left(\rho_f + \frac{N m_e}{\volm}\right).
$$
and the inertial mass is then also a scalar,
\begin{equation}
  \label{massi}
  M= \rho_m \volm.
\end{equation}
Before presenting the results for $\volm$, in the following section we
prove that the inertial mass,  derived from Eq. \ref{u0} is consistent
with that appearing in  the equipartition of its translational kinetic
energy.

\subsubsection{Mass from equipartition}
The  equipartition  theorem  applied  to the  multiblob  translational
energy  $\ang{\bU_0^2} =  \kt/M_{\tex{th}}$, provides  an alternative
route  to its mass which we have called ``thermal mass'' $M_{\tex{th}}$. 
In the following calculation for $M_{\tex{th}}$
we consider a  neutrally buoyant particle in  a fluid
with  density  $\rho_f$, so that $\volm=M_{\tex{th}}/\rho_f$.

We start by quoting the equipartition of the fluid degree's of freedom, which
can be generally written as \cite{Balboa_tesis}
\begin{equation}
  \label{ef}
  \langle v^{\alpha}(\br) v^{\beta}(\br') \rangle = \frac{\kt}{\rho_f} \bP^{\alpha,\beta} \delta(\br-\br'),
\end{equation}
where $\alpha$ and $\beta$ are the components of the covariance (dyadic).
Recall that in the compressible formulation $\bP^{\alpha,\beta}=\delta_{\alpha,\beta}$, however, in the incompressible fluid, fluctuations along different directions are correlated (within the same cell). 
The particle translational velocity is,
\eqn  
\bU_0 =\frac{1}{N} \sum_{i}^N \bJ_{i} \bv = \Jm \bv
\eqnend
Using the no-slip condition over each blob $\bu_i=\J_i\bv = \int \delta_h(\br-\bq_i) \bv(\br) d\br$ one gets,
\begin{equation}
  \langle U_0^{\alpha} U_0^{\beta}\rangle = \frac{1}{N^2}
  \sum_{i,j}  \int \delta_h(\br-\bq_i)  \langle  v^{\alpha}(\br) v^{\beta}(\br') \rangle
  \delta_h(\br'-\bq_j)  d\br d\br' 
\end{equation}
In this evaluation we are implicitly assuming that 
that the particle translational degrees of  freedom are much slower 
than the fluid velocity so terms involving $\left\{\bq_i\right\}$ can
be frozen in the fluid thermal average (as customarily done in colloids,
due to the very large Schmidt number)
Using also Eq. \ref{ef},
\begin{eqnarray}
  \langle U_0^{\alpha} U_0^{\beta}\rangle &=& \frac{\kt}{N^2 \rho_f}
  \sum_{i,j} \int \int  \delta_h(\br-\bq_i) \bP^{\alpha,\beta} \delta(\br-\br') \delta_h(\br'-\bq_j)  d\br d\br' =\\
  \nonumber
  &=&  \frac{\kt}{N^2\rho_f} 
  \sum_{i,j} \int \delta_h(\br-\bq_i) \bP^{\alpha,\beta} \delta_h(\br-\bq_j)  d\br =  \frac{\kt}{\rho_f} \frac{1}{N^2} \sum_{i,j} \J_i \bP^{\alpha,\beta} \S_j =\\
  \nonumber
  &=& \frac{\kt}{\rho_f} \Jm\bP^{\alpha,\beta} \Sm
\end{eqnarray}
Thus the thermal mass of the multiblob coincides with its inertial mass matrix.
In case of an isotropic body $\Jm\bP \Sm = \volm^{-1} {\bs 1}$ so
$M_{\tex{th}}=\rho_f \volm$ in agreement with Eq. \ref{massi}.
The equipartition of translational kinetic energy is recovered as,
\begin{equation}
  \label{eb}
  \langle U_0^2\rangle = \frac{\kt}{M}
\end{equation}

\subsection{Inertia tensor}

\subsubsection{From rotational inertia} 
Equation  (\ref{rot1b}) can  be applied  to multiblobs  with
arbitrary shape and mass distribution. It permits to derive the inertia tensor by evaluating
the initial angular acceleration upon the application of a given torque ${\bs \tau}$ at $t=0$.
In particular, we consider a fluid at rest and a particle with ${\bs \Omega}=0$, which upon application of
${\bs \tau}$ will increase its angular velocity according to  $\dot {\bs \Omega} = \mIt^{-1} {\bs \tau}$.
From Eq. (\ref{rotnb}), the resulting tensor of inertia of the neutrally buoyant particle is,
\begin{equation}
  \label{itnb}
  \mIt = \rho_f \mI \mIc^{-1} \mI
\end{equation}
For isotropic bodies (like a sphere, the icosahedron or any Platonic solid) 
both $\mI$, $\mIc$ and also $\mIt$ are scalars in the continuum setup (diagonal matrices with three similar eigenvalues)
Again, in the discrete setup, $\mIt$ is an operator which generally depends on the grid coordinate and the
rotation direction. We advance that in the discrete grid, deviations of the constant scalar ``behavior'' of $\mIt$ 
were found to be small (about $1\%$) as observed with the particle volume.

\subsubsection{From equipartition} 
The inertia tensor can be  also obtained  from the  equipartition of  the  rotational energy.
In the case of neutrally buoyant particles we use the angular momentum
defined in Eq. \ref{angularmom} and evaluate its covariance. We first have $\bs{L}=\rho_f \mI \bs{\Omega}$
so,
\begin{equation}
  \langle \bfL \bfL^*\rangle = \rho_f^2 \mI \langle {\bs \Omega} {\bs \Omega}^*\rangle  \mI^t,
\end{equation}
and from the definition (\ref{angularmom}) one gets,
\begin{equation}
  \langle \bfL \bfL^*\rangle =  \rho_f^2 \langle \sum_{i,j} \left(\bfs_i\times \bJ_i \bv \right)\, \left(\bfs_j \times \bv^* \bS_j\right)\rangle
 = \kt \rho_f \, \mIc.
\end{equation}
which  can  be  shown by  using  using  Levi  Civita tensors  for  the
cross-products,  as  in  Eq.   (\ref{mict})  and  also  that  $\langle
\bv({\bf r}) \bv^*({\bf r'})  \rangle = (k_B T/\rho_f) \bP \delta({\bf
  r}-{\bf  r'})$. Combining both results, one  gets the  consistent  equipartition
relation,
\begin{equation}
\label{equiang}
  \langle {\bs \Omega} {\bs \Omega}^*\rangle = \kt\, \mIt^{-1}
\end{equation}
Numerical and analytic results for $\mIt$ are presented in the next section. We advance
that numerical values of $\mIt$ obtained from the inertial route (angular acceleration after an external torque) and from equipartition (using Eq. \ref{equiang}) agree within less than about $1\%$ for any value of $d$ considered.

\subsection{Analytic and numerical evaluation of mass, inertia moment and rotation radius}

\subsubsection{Mass and volume} 
\begin{figure}
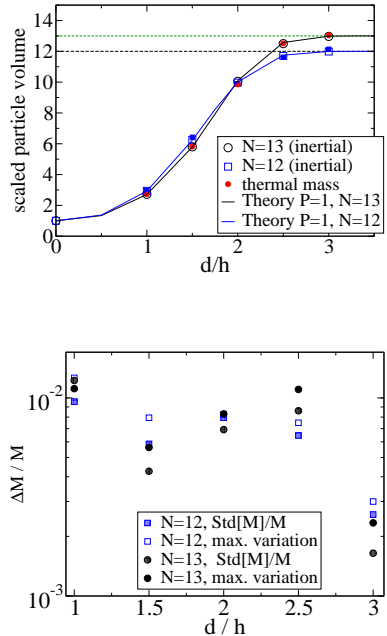

  \includegraphics[scale=0.2]{mass_inertia_and_hydrovol.eps}\\
  \vspace{1.0cm}
  \includegraphics[scale=0.2]{STD_MASS2.eps}
  \caption{\label{fig:mass}  Top:  Volume  of the  multiblob  particle
    scaled  with the  isolated  blob volume  $\volm/\volt$ versus  the
    neighbor vertex distance $d$  of the icosahedron model. Numerical
    results  were carried out  in an  incompressible fluid  (where the
    single blob  volume is  $\volt=(3/2) \vol$ due  to the  added mass
    effect).  Filled  symbols corresponds to  thermal measurements and
    empty  symbols  come from the inertial   evaluation  of  the  particle  mass
    $M=\rho_m \volm$ in  the neutrally buoyant case ($\rho_m=\rho_f$).
    Solid   line   corresponds    to   the   theoretical   result   of
    Eq.  (\ref{mass2}) for  the compressible  ($\bP=1$) case:  in this
    case the  mass is  scaled with the  blob mass $\rho_f  \vol$, with
    $\vol=8\,h^3$. Relative maximum variation of the particle mass (or volume $\volm$)
    over the grid versus the vertex distance $d$.}
\end{figure}

The mass of the multiblob can be also expressed as,
\begin{equation}
  \label{mass2}
  M= \rho_f \volt_m + Nm_e= \rhom \volm=\frac{N\rhom \volt}{1+ (N-1) \widehat{\alpha}}
\end{equation}
where the non-dimensional factor $\widehat{\alpha}$ measures  the average  overlap per  blob,
\begin{equation}
  \label{alphahati}
  \widehat{\alpha}= \frac{1}{N(N-1)}\sum_{i,j\ne i} \langle \J_i \bP \S_j \rangle_{grid} \tilde{\vol},
\end{equation}
and involves the average of $\left(\J_i \bP \S_j {\bs e} \right) \cdot
{\bs  e}$ acting over  all possible  unit vectors  ${\bs e}$  and grid
locations; we  noted it as $\langle  ...\rangle_{grid}$.  This overlap
factor  $\widehat{\alpha}$   ranges  from  $\widehat{\alpha}=1$  (full
overlap between blobs, or $d=0h$) to $\widehat{\alpha}=0$ (no-overlap,
taking place at  large $d$).  In the first case,  the particle mass is
equivalent  to a  single  blob  with excess  mass  $Nm_e$ and  density
$\rhom=\rho_f+Nm_e/\volt$, while  in the former,  the particle density
is equal  to that of  single blob with  mass $m_e$. In  principle, the
zero  overlap  limit  ($\J_i\bP\S_j   =0$)  is  only  reached  in  the
compressible case  ($\bP={\bf 1}$)  where $\J_i\bP \S_j>0$  and equals
zero when the two blobs are at distance $d>3\,h$ along one direction (using 3pt kernels).  
In the incompressible case the
dependence of $\widehat{\alpha}$ with  the interblob separation $d$ is
not  that   trivial  because  $\bP$  could   induce  negative  overlap
contributions  $\J_i\bP  \S_j<0$. Yet,  the  ``dynamic''  mass of  the
multiblob in a 3D incompressible fluid should consistently contain the
fluid  added mass.   In other  words  if $\rhom  \mathbb{V}_m$ is  the
multiblob mass in the compressible $\bP=1$ case; in the incompressible
fluid it  should be  $(3/2) \rhom \mathbb{V}_m$.   This is true  for a
single blob $\volt=(3/2)\vol$  but it is not trivial  in the multiblob
due to  the effect of $\bP$  acting on a spread field made up of 
significant  kernel overlap. For
neutrally buoyant  multiblobs the added mass  consistency implies that
$\volm/\volt  =  \mathbb{V}_m/\vol$.  It is  possible  to
analytically  evaluate  the   multiblob  volume  $\mathbb{V}_m^{-1}  =
(1/N^2) \sum_{i,j} \J_i \S_j$ 
in the $\bP=1$ case. The overlap integrals are just $\J_i \S_j =
\int \delta_h({\bf r}-{\bf q}_i) \delta_h({\bf r}-{\bf q}_j) d{\bf r}$.
We  have performed  such calculation  in the  continuum  setting using
$\Phi(x)=(2/3)\cos(2\pi  x/3)^2$  (which  is  extremely  close  to  3pt
Peskin's function).  Theoretical results for  $\mathbb{V}_m/\vol$ shown
in Fig. \ref{fig:mass} are in perfect agreement with $\volm/\volt$ obtained
from  the  incompressible solver,  indicating that  the
significant blob-overlaps do not alter
the consistency of the added mass  effect of  the  multiblob.

Figure \ref{fig:mass}  presents results for the ``filled''  $N=13$  icosahedron and
for the icosahedron ``shell'' with $N=12$.  Numerically, the  volume  $\volm$ 
was obtained from the mass of neutrally  buoyant particles $\volm=M/\rho_f$
which was evaluated following both inertial and  thermal routes.  In  the thermal
route  we  measured  $M_{\tex{th}}=\kt/\ang{U_0^2}$  averaging  the  squared
particle   velocity   over   long   simulations   where   hydrodynamic
fluctuations  where set  in the  fluid  at finite temperature.  We  also
measured $M_{\tex{th}}$ from the initial ballistic regime of the mean square
displacement of  the particle (see  below), getting the same outcome.
Figure \ref{fig:mass}  shows the consistency between  the inertial and
thermal measurements.  As shown in the same figure, relative mass
variations $\Delta M/M  = \Delta \volm/\volm$ are found  to be smaller
or about than  $1\%$ for any vertex separation  $d$ considered.  It is
somewhat surprising that for $d>1.5h$ relative variations in multiblob
volume  (or mass)  are significantly  smaller  than that  found for a
single blob.

\subsubsection{Moment of inertia and rotation radius}
\label{sec:rot}
We performed evaluations of the moment of inertia of the multiblob
for different $d$, both from the inertial and thermal routes. Both
methods provided perfectly consistent results. The deviation of the 
moment of inertia
over the mesh was found to be quite small, typically about $1\%$ or less.
The moment of inertia of the multiblob and its mass can be used to determine 
a physical radius of the multiblob. The moment of inertia of a solid icosahedron
measured with its radius $a$, equals that of the sphere. In fact, instead of presenting results for $\mIt$,
we find more practical to present the results for this ``radius of inertia'' $R_I$,
defined from the relation,
\begin{equation}
  \mIt = \frac{2}{5} M R_I^2 
\end{equation}
It is  important to stress that,  like the moment  of inertia $\mIt$
and the mass  $M$, this radius $R_I$ is  a ``geometrical'' property of
the  multiblob shape and  the interpolator  $S$ used.  This contrasts
with the  hydrodynamic radius and other measurements  of the multiblob
size  presented  below which  depends  on  how  the object  physically
perturbs   the   fluid    (which   therefore   implies   finite   size
effects). Table \ref{tab:rad} presents  results for $R_I$ obtained for
different icosahedron sizes in  the incompressible setting. As stated,
$\mIt$ obtained  from equipartition [Eq. (\ref{equiang}]  and from the
relation  between  angular acceleration  and  external  torque are  in
perfect agreement (about $1\%$  deviation).  On the other hand, unlike
the  multiblob  scaled  volume,  the  analytic  calculation  of  the
moment of  inertia from the expression  (\ref{itnb}) (using $\bP=1$,
compressible fluid) reveals that  $R_I$ differ in the compressible and
the incompressible settings.  For  small blob overlaps ($d>3\,h$) both
(compressible and incompressible) values of $R_I$ converge, however as
the overlap increases ($d\leq 1.5\,h$) the solenoidal projection $\bP$
of the inner multiblob spread  field makes $R_I^2$ about $2/3$ smaller
in the incompressible case.

\section{Multiblob Stokeslet behavior}
\label{sec:stokes}
As a  first step in the  analysis of the hydrodynamic response of the
multiblob model,  we analyze  its Stokeslet from
the hydrodynamic  perturbation it creates  at long distances  from its
center. Like in subsequent analyses we will derive effective hydrodynamic radius of
the multiblob particle by comparing its behavior with that of a rigid
sphere. We will then study  the dynamic  response of  the particle  from  its mean
square  displacement  and calculate  its  self-diffusion constant,  to
cross-check with  the friction  coefficient via the  Einstein relation
\cite{BalboaStokesEinstein}.

\subsection{Hydrodynamic radius}

To  measure the particle  hydrodynamic radius  $R_H$ 
of the multiblob we ascribe the traction  exerted by its  own perturbative
field \cite{Maxey1983,Duenweg2009,Balboa2012a}  to the friction  a rigid (no-slip)
sphere would  feel under the  same circumstance, $6\pi \eta  R_H u_0$. According to the Stokes relation, 
the friction coefficient $6\pi \eta  R_H$ is just the ratio  between   a  (small)  pulling   force  $F_0$  and   the  particle
translational velocity, $U_0$.  In a  periodic box of size $L$, 
the  drag  increases roughly with $L^{-1}$ due to  the
interaction of the particle with its own periodic images. 
This is a real physical  effect which corresponds to the  increase of
drag felt by a particle in a regular array of particles separated by  distance $L$.
The analytic solution of this problem (Hasimoto   \cite{Hasimoto})
permits to extrapolate the hydrodynamic radius to infinite boxes $L\rightarrow
\infty$, the   dominant term in the drag reduction scaling like $1/L$.
Assuming that the Stokes relation is valid for a single particle 
in our periodic box, its drag $F= 6\pi\eta R_H(L) u_0(L)$
is independent on the box size, leading to the following 
ratio between $R_H$ in two limiting box sizes \cite{Duenweg2009,Balboa2012a}
\begin{eqnarray}
  \label{rhinf}
  \frac{R_H(\infty)}{R_H(L)} &=& 1 - 2.84 \frac{R_H({\infty})}{L} + \mathcal{O}\pare{\left(\frac{R_H({\infty})}{L}\right)^3} ; \,\mathrm{i.e.,}\\
  R_H(\infty) &=& \frac{R_H(L)}{1 + 2.84 \frac{R_H(L)}{L}},
\end{eqnarray}
which  provides $R_H(\infty)$  from the  outcome any  finite  box $L$.
Unless explicitly  stated, we  use $R_H$ to  indicate $R_H(L\rightarrow
\infty)$.   Figure  \ref{fig:rh} shows  $R_H$  versus the  icosahedron
vertex  distance  $d$.  The  first  observation  is  that both  models
($N=12$ blobs in an empty  icosahedron and the filled icosahedron with
$N=13$)  have  quite similar  hydrodynamic  radius;  the $N=12$  model
having a  slightly larger $R_H$  at the largest $d$  considered.  This
indicates that  the monopole contribution to  the particle disturbance
at long  distances is essentially  controlled by the ``shell''  of the
multiblob and not by its core or inner part.  The maximum variation of
$R_H$ along the grid (in Fig.  \ref{fig:rh} bottom) was found to be of
the same order than  for a single blob, and even smaller  for $d > 1.5
\,h$.   Thus  kernel  overlaps  in  the  multiblob  are  not  strongly
affecting  the  grid  dependence   on  $R_H$.   The  highest  relative
variation  of $R_H$  over  the  grid were  found  for $d=1\,h$  ($5\%$
variation) and $d=1.5\,h$  ($2.8\%$) and $\Delta R_H/R_H$ sustantially
decreases for  $d>1.5h$ (see  Fig. \ref{fig:rh}). 

Figure \ref{fig:rh}  compares $R_H$ with  other relevant sizes  of the
model: its radius (i.e., the  radius of the embedding sphere) $a\simeq
0.95106\, d$  and the Fax\'en radius  of the average  kernel, given by
its  second  moment  $\Jm[(\br-\bq)^2]^{1/2}$ [see Appendix, Eq. (\ref{jm2})],
whose discussion is deferred to Sec. \ref{sec:consistency}.

\subsection{Self mobility}

\begin{figure}
  \centering
  \includegraphics[width=0.7\textwidth]{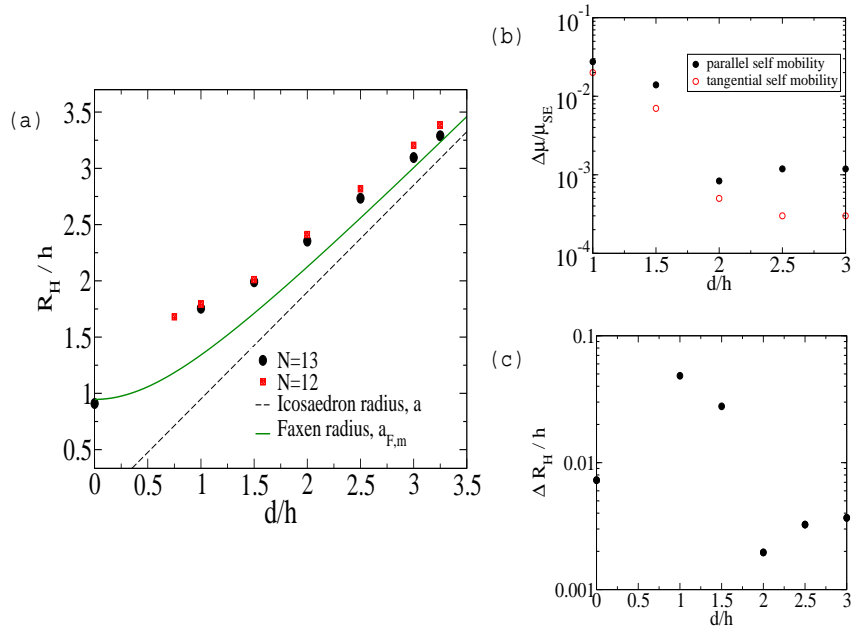}
  \caption{(a) Hydrodynamic  radius  in infinite  box
    extrapolated  from  simulations  with  $L=64\,h$  and  $L=128  h$.
    (b) Relative deviation of the self mobility tensor components from the isotropic
    case (deviations scaled with the parallel Stokes-Einstein mobility) 
    (c) Maximum variation of  $R_H$ over the grid. (b) and (c) corresponds to $N=13$.
    \label{fig:rh}
  }
\end{figure}

Pulling experiments of single particles through the fluid grid
also serves to evaluate its self mobility tensor $\bs{\mu}$ via the relation,
\begin{equation}
  \bU_0 = {\bs \mu} \bF_0
\end{equation}
The velocity of the particle in the pulling direction ${\bf n}$ and
in  the  perpendicular  direction  ${\bf  t}$  is  determined  by  the
components  of  the  mobility  $\mu_{n} ={\bs  \mu}\cdot{\bf  n}$  and
$\mu_{t}  ={\bs \mu}\cdot{\bf  t}$.  The  self mobility  tensor  of an
isotropic particle,  with three perpendicular axis of  symmetry, is an
scalar  ${\bs \mu}=\mu  {\bf 1}$  \cite{Happel-Brenner-book}.   Ideally this
should be the case of our icosahedron model, with a parallel mobility
$\mu_{n}=\mu_{S}$   (where  $\mu_{S}=1/(6\pi   \eta   R_H)$  is   the
Stokes mobility) and a vanishing lift term $\mu_{t}=0$.  The
maximum relative deviations  $\Delta \mu/\mu_{S}$ over different grid
locations  are shown  in  Fig.  \ref{fig:rh}  for different  particle
sizes  $d$.  Deviations  are less  than  $1\%$ and  decrease to  about
$0.1\%$ for $d>1.5 h$.  In  the smallest particles ($d=h$) these $1\%$
deviations in tangential mobility  around zero (average value over the
grid) induces a  slight oscillation in the particle  translation and a
slight ``heading''  in its motion due to  (similarly small) variations
in the translational-rotational mobility  over the grid (which is also
zero for isotropic particles \cite{Happel-Brenner-book}).

\subsection{Mean square displacement}
\begin{figure}
  \centering
  \includegraphics[width=0.4\textwidth, angle = 270]{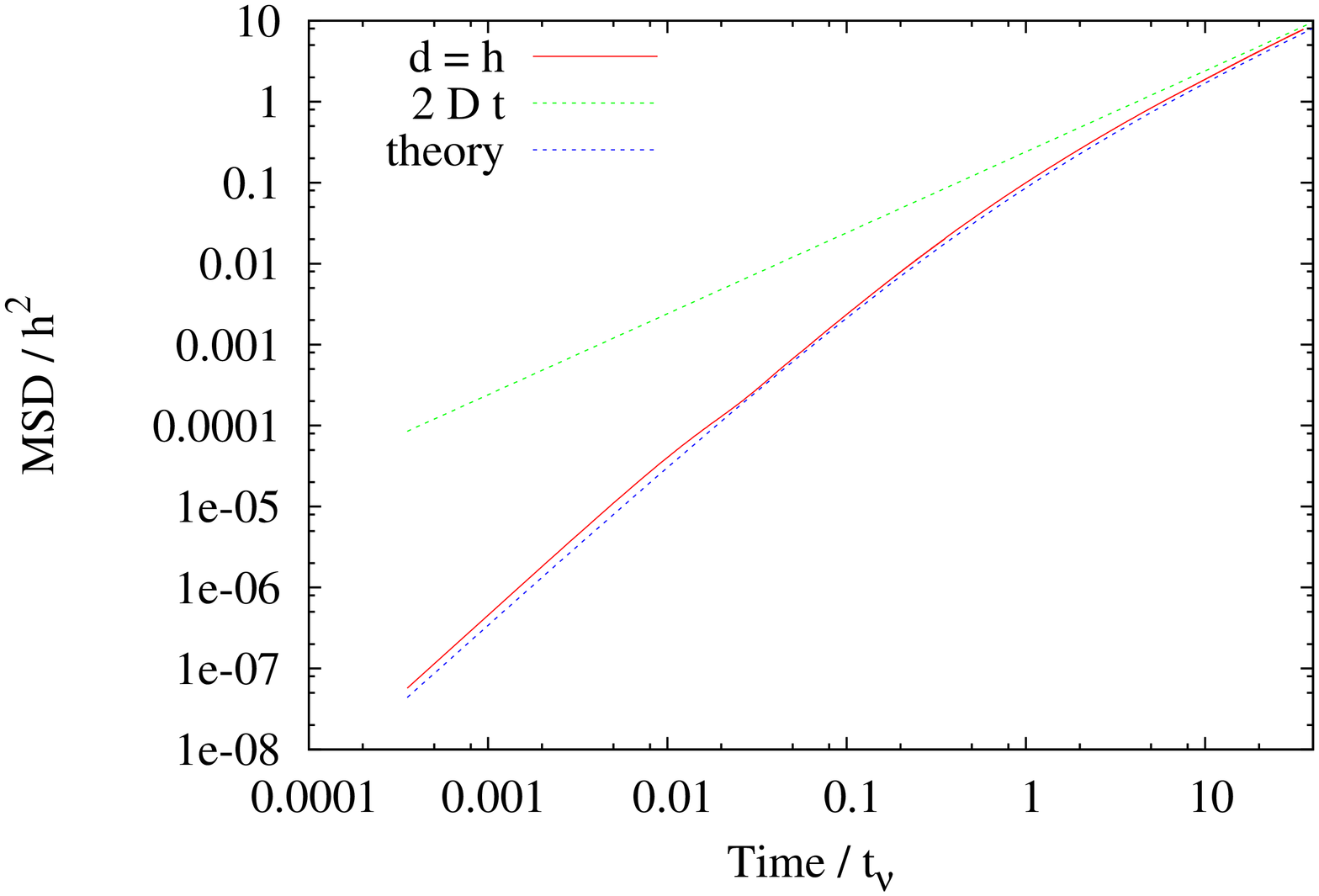}
  \includegraphics[width=0.4\textwidth, angle = 270]{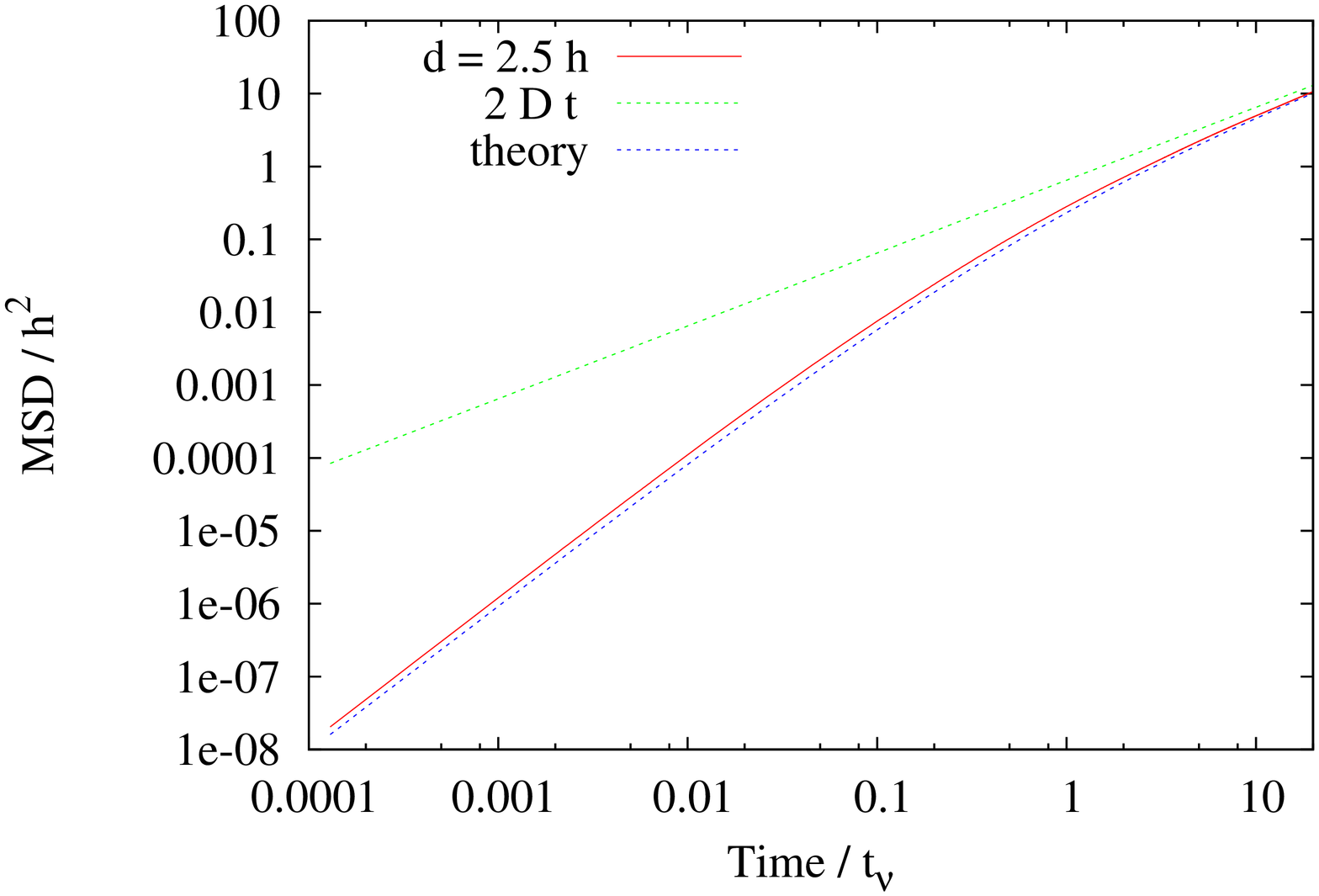}
  \caption{Mean square  displacement of multiblob  icosahedrons 
    compared with the analytic result for a sphere with the same
    hydrodynamic radius.  Results correspond to the filled icosahedron $N=13$ in $L=64\,h$ boxes
and other data in Table \ref{tab1}. The fluid temperature is  $k_B\,T=3$ and Schmidt
    number $Sc=40$. The diffusion constant $D$ comes from the Einstein relation,
$D=\kt/(6\pi\eta R_H)$.
    \label{MSD}}
\end{figure}

\begin{figure}
  \centering
  \includegraphics[width=0.4\textwidth, angle = 270]{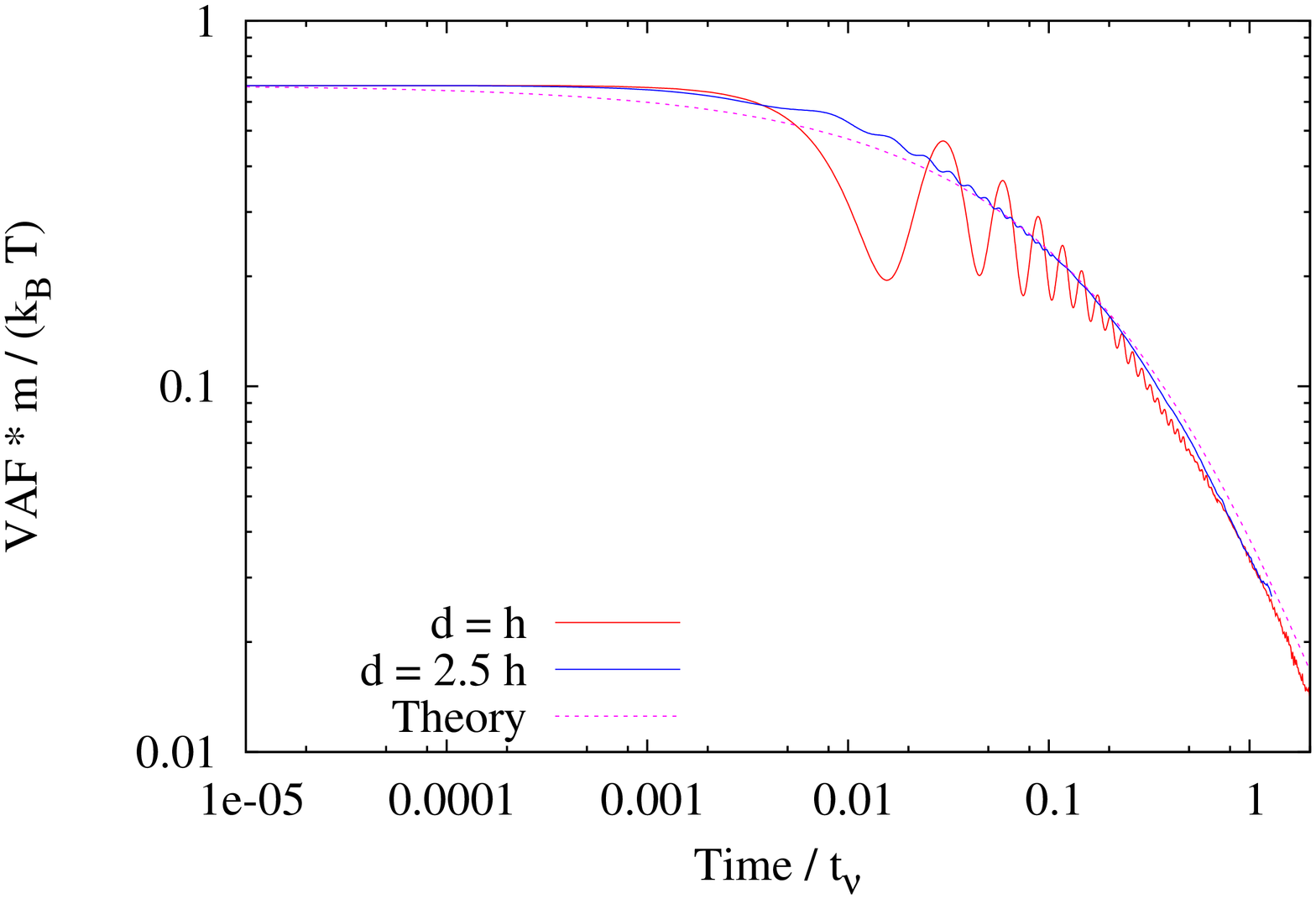}\\
  \includegraphics[width=0.4\textwidth, angle = 270]{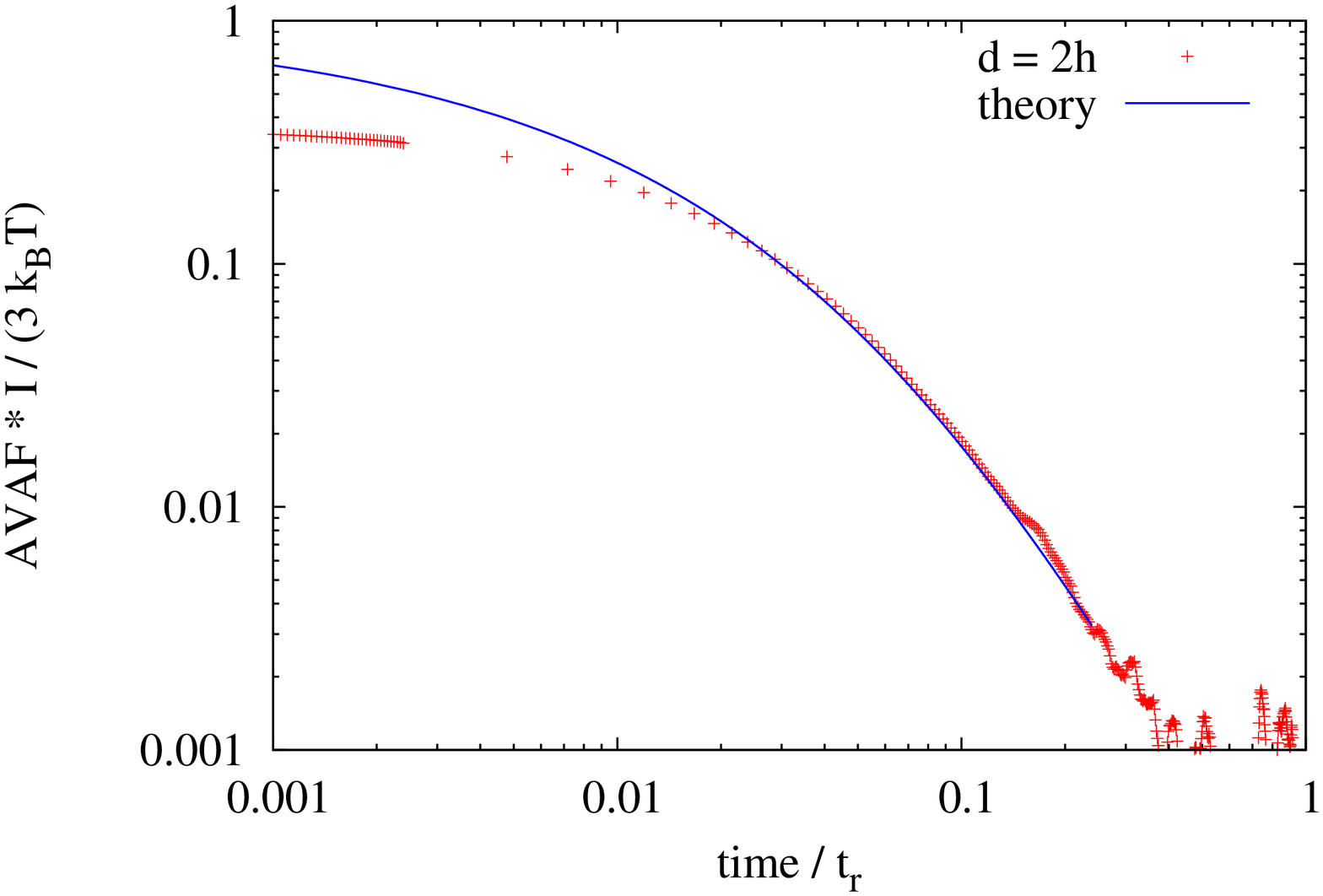}
  \caption{\label{VAF} Top panel: Velocity  autocorrelation function  for the  MSD  curves of
    Fig. \ref{MSD}. Time is scaled with $t_{\nu}=R_H(L)^2/\nu$
    and the theoretical result corresponds to a sphere with
    radius similar to the hydrodynamic radius of the  icosahedron  in an
    incompressible fluid. Bottom panel shows the time autocorrelation of the angular velocity $\langle \Omega(t)\Omega(0)\rangle$
for $d=2\,h$ and compare it with the theoretical result for a sphere \cite{Zatovsky1969} of radius $R_I$.
The time axis is scaled with the rotation relaxation time $\tau_r=8\pi\eta R_I^3/(6 \kt)$ and the y-axis with the sphere $\langle \Omega^2\rangle $ covariance: $3\kt/I_{sp}$ with $I_{sp}=2/5 M_{sp} R_I^2$ and $M_{sp}=(4/3)\pi \rho_p R_I^3$. All cases correspond to neutrally buoyant particles, $\rho_p=\rho_f$.}
\end{figure}

We  have analyzed  the Brownian  motion of  the  multiblob icosahedron
including  hydrodynamic  fluctuations in  a  fluid  with zero  average
velocity.  The  objective is to  compare its mean  square displacement
(MSD) with  that of  a sphere so  as to  check the consistency  of the
model  at several  scale  times.  In particular,  at  small times  the
ballistic  regime depends  on the  particle mass.  At long diffusive times,
the self friction $\xi$ and the  diffusion coefficient  are  related $D=kT/\xi$, according to the Einstein relation.
At large  enough Schmidt number $\mathrm{Sc}=\eta/(\rho_f D)$ (typically for
$\mathrm{Sc}>20$)  the friction  coefficient converges  to  the Stokes
value $\xi=6\pi\eta  R_H$, however at  small $\mathrm{Sc}$ deviations
from  the Stokes limit  occur (see  Ref. \cite{BalboaStokesEinstein}).   We have
observed  this  effect  in  simulations with  $\mathrm{Sc}\simeq  10$,
however the  results presented hereafter correspond  to $\mathrm{Sc}\simeq 40$
for   which   the   Stokes   friction  holds   with   large   accuracy
\cite{BalboaStokesEinstein}.   In  figure  \ref{MSD}  we  draw the  MSD  of  two
differently sized icosahedrons,  $d= h$ and $d =  2.5\,h$ in a $L=64\,h$
box and compare  the result with the theoretical  MSD for spheres with
the same hydrodynamic radius and mass in an incompressible fluid with
the same density and viscosity. The theoretical MSD is calculated from
the  friction  memory  function   derived  by  Bedeaux  and  Mazur  in
\cite{Mazur1974}.   Both cases  agree quite  well  with the
theory for spheres over all  time scales.  In particular, the particle
(thermal) mass determined by the  slope of  the ballistic  regime equals
that obtained by other  means (equipartition and inertial methods)
and at longer times, the diffusion constant is consistent with the observed friction
(Einstein relation).  Slight deviations  at  times
smaller  than  the viscous  time  $t_\nu$ are  visible  for  the $d  =
h$. These  come out from the  coupling of the high  frequencies of the
elastic structure and the translational motion of the multiblob.  Such
coupling is  not a  numerical artifact, but  rather a  physical effect
which   has   been  recently   theoretically   studied  by   Felderhof
\cite{Felderhof2014}.   The effect  is  clearly observed  in the  velocity
autocorrelation function (VAF) shown in Fig. \ref{VAF} and consists on
a  jitter of  the  translational velocity  of  the elastic  structures
moving in incompressible fluid.  Consistent with the ratio between the
spring  time $t_{sp}=(\rho_f  \volt/k_{sp})^{1/2}$  and the  momentum
diffusion  time  $t_{\nu}=R_H^2/\nu$  (we  used  $t_{\nu}/t_{sp}
\simeq  80$),  the  amplitude  of  such  oscillations  start  at  about
$0.01\,t_{\nu}$ and decay (probably exponentially \cite{Felderhof2014}) in
time.  Although  we have  not yet studied  this in detail,  we observe
that  the  jitter is  drastically  reduced  if  the particle  size  is
increased  (see Fig.  \ref{VAF} for  $d=2,5\,h$).  This  is consistent
with the analysis presented in  Ref. \cite{Felderhof2014} and also with the
decrease   in  the   non-dimensional  group   $(kT/k_{sp}  a^2)^{1/2}$
introduced before.  Contrary to the prediction of Ref. \cite{Felderhof2014}
we do not  observe the suppression of the added mass  effect in the VAF
of the elastic  structure, whose value at $t=0$ is  still reduced by a
factor $2/3$  (see Ref. \cite{Balboa2014,Balboa_tesis}).  Note that in the  blob model
the particle is  ``filled'' with fluid and thus  conserves its volume,
which is not true for the theoretical framework \cite{Felderhof2014}. A study
of  this effect should  probably require  working in  the compressible
formulation.

\begin{table}[t!]
  \begin{tabular}{|l | c |}
    \hline
    Parameter & Value \\
    \hline
    \hline
    Fluid density $\rho_f$ & $5.0$\\
    \hline
    Dynamic viscosity $\eta$ & $2.5$\\
    \hline
    Box size $L$ & $128.0$ or $256.0$ for icosahedrons \\
    & $64.0$  for blobs\\
    \hline
    Cell size $h$ & 2.0\\
    \hline
    Vertexes distance $d$ & Variable, from $h$ to $3.25 h$\\
    \hline
    Interblob spring constant $k_{sp}$ & $2\times 10^4$ \\
    \hline
  \end{tabular}
  \caption{Common parameters used in all simulations. All particles 
    neutrally buoyant $m_e=0$}
  \label{tab1}
\end{table}

\section{Rotlet and rotational diffusion}
The  angular friction on  a rotating  sphere $8\pi  \eta R^3$  and the
sphere rotational diffusion are  also related by the Einstein formula,
$D_r=  k_BT  /(8  \pi\eta  R^3)$.  We have  evaluated  the  rotational
diffusion  $D_r$ of  multiblobs  with different  sizes  from the  time
integral of the autocorrelation  of the multiblob angular velocity. As
for the  hydrodynamic radius in  relation with the Stokeslet,  we can
evaluate  an  effective multiblob  radius  associated  to its  rotlet,
i.e. to its inertialess rotation  dynamics.  This is just, $R_r \equiv
\left[k_BT /(8 \pi\eta D_r)\right]^{1/3}$.  The effect of the particle
rotation on the  fluid propagates like $1/r^3$ so  finite size effects
are  less relevant  than in  the Stokeslet  case.  As  shown  in Table
\ref{tab:rad}, for any $d$,  the ``rotlet''radius $R_r$ is consistent
with  $R_I$ and $R_H$, with a particularly good agreement in the range $1.5\leq d/h \le 2.5$.

Figure \ref{VAF}(bottom) shows the time autocorrelation of the angular
velocity  and compares  our numerical  results for  $d=2\,h$  with the
theoretical curve derived in Ref.  \cite{Zatovsky1969} for a sphere of radius $R_I$.  
The time-axis is scaled with the  rotational relaxation time $\tau_r=1/(6D_r)$  and the y-axis
with  $3k_BT/I_{sp}$  with  $I_{sp}=(8/15)\pi\rho R_I^5$. This scaling
permits to observe deviations from the sphere  behaviour in different regimes.
In particular, deviations are expected at  small  times  ($t<0.01
\tau_r$) because of the difference in the inertial masses (the mass of
sphere  $4\pi\rho  R^3/3$,  differs  from  the
multiblob  mass in  Fig.   \ref{fig:mass}).  Interestingly, as the  momentum
diffusion takes over, both  (theoretical and numerical) curves overlap
quite exactly and  converge to the limitting $t^{-5/2}$  power law.  
The excellent  agreement found in diffusive regimes
using $R_I$ as the scaling radius explains
the  slight  difference between  $R_I$  and  $R_r$;  as the  later  is
obtained     from      the     whole     integral      of     $\langle
\Omega(t)\Omega(0)\rangle$,   which  also   includes   the  transition
(inertial-to-diffusive) regime into account.

\section{Velocity profiles}
\label{sec:vel}

Over the  following sections  we focus on  the hydrodynamics  at close
distances from  the particle center $r  \sim R_H$.  We  start with the
fluid velocity  profile past a  multiblob particle moving  at constant
velocity $\bU_0$ in a fluid which would be otherwise at rest.

\subsection{Empty and filled model comparison}
\begin{figure}
  \centering
  \includegraphics[width=0.5\textwidth,angle=270]{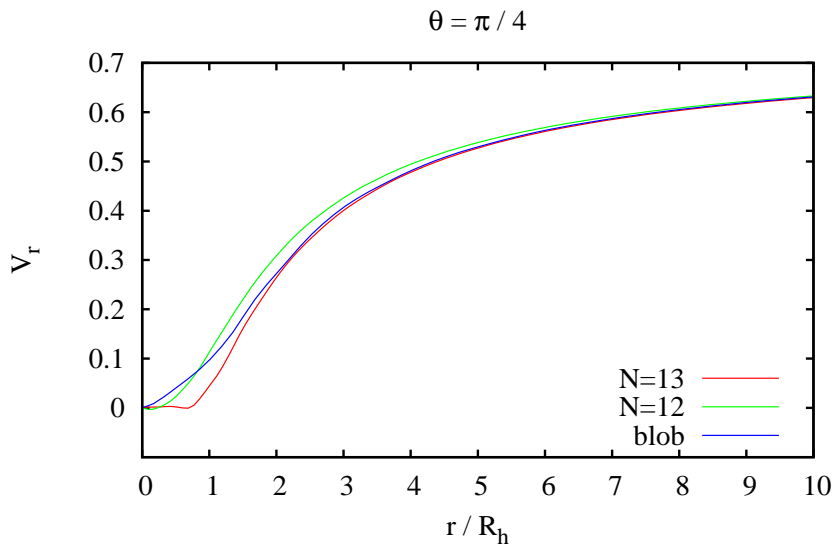}
  \caption{Radial  component   of  the  velocity   profile  along  the
    direction forming $\theta=45^o$ with the direction of the constant
    plug  velocity field.  Results  corresponds to  the zero  Reynolds
    limit for a blob particle ($R_H(L) =  1.86\,h$ in a $L = 64\ h$ box),
    and a multiblob  particle with  $N=12$ (hydrodynamic radius  $R_H(L) =
    1.83\,h$) and $N=13$ (filled icosahedron with $R_H(L)=1.83\,h$) in both
    cases with $d = h$ and boxes of $L = 128\ h$.
    \label{fig:velcomp}
  }
\end{figure}

Figure \ref{fig:velcomp} compares the velocity profile past the $N=12$
empty icosahedron and  the filled one with $N=13$  blobs.  The result
for the single  blob is added for reference.  In  these test the ratio
$L/R_H(L)$  is  approximately  fixed   to  have  similar  finite  size
contributions.    The    first   feature   to    highlight   in   Fig.
\ref{fig:velcomp},  is that  the empty  multiblob $N=12$  suffers from
substantial fluid  leakage inside its  body kernel. In this  sense its
behavior resembles the single blob  case. By contrast, the extra blob
placed at  the icosahedron center ($N=13$ multiblob)  avoids the fluid
leakage and furnishes  a particle with an impenetrable  core where the
relative  particle-fluid  velocity  is  zero. The  filled  icosahedron
multiblob,  with $N=13$,  seems therefore better  suited to  simulate
suspensions of rigid impermeable colloids and in the remainder of
this section we focus on the near-field behavior of the $N=13$ model.

\subsection{Filled multiblob and sphere}
\begin{figure}
  \centering
  \includegraphics[width=0.5\textwidth]{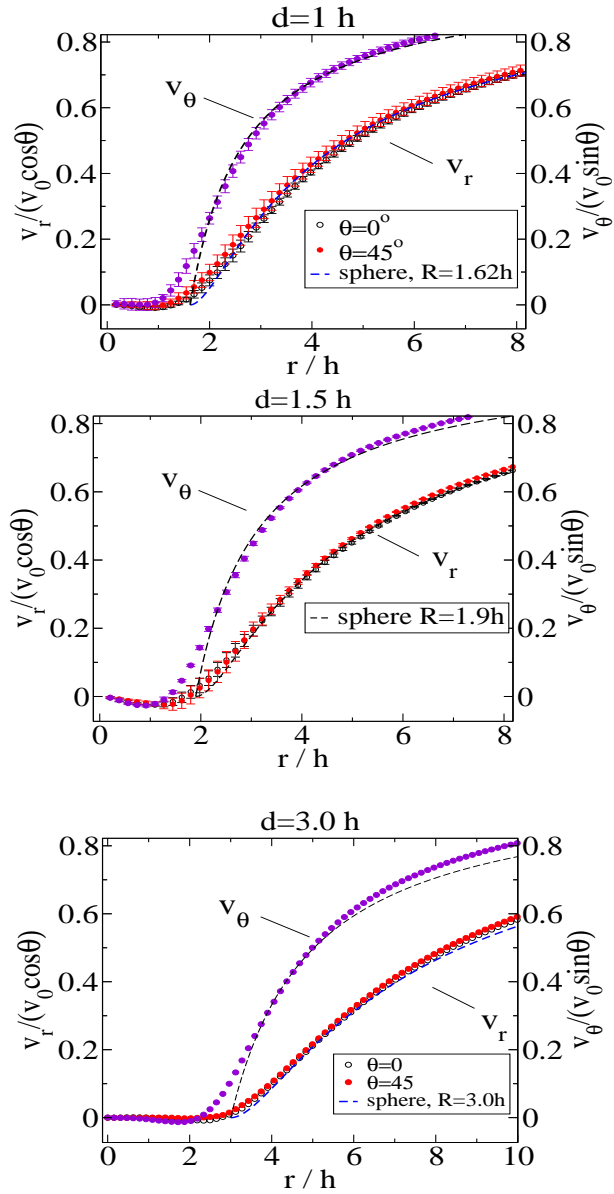}
  \caption{
    \label{fig:vel}
    Normal and  tangential components of  the velocity profiles
    past  a  multiblob   particle  (filled  icosahedrons  with  $N=13$)
    moving  at  constant velocity  in  the  zero Reynolds  limit.
    Results  correspond  to several  values  of  the vertex  distance
    $d$. Dashed lines  inidicate the velocity profile  of a rigid
    sphere with an effective ``no-slip'' radius $R_{s}$, specified for
    each case. All cases corresponds to $L=128h$ (with $h=2$).}
\end{figure}

In the following sections  the hydrodynamic behavior of the multiblob
particle is analyzed by comparison  with sphere results.  The sphere is
used  here as  the  reference model,  in  part due  to  the wealth  of
theoretical (and experimental) results  it offers.  The fluid velocity
past multiblob particles (in the  zero Reynolds regime) is compared in
Fig.  \ref{fig:vel}  with the analytic  solution for a  rigid sphere
with  no-slip boundaries.   The best  fit  to the  sphere provides  an
estimation of  the estimation  of the location  of its  effective {\em
  no-slip surface}, related to the its hydrodynamic size {\em at close
  distances}.  Due to the  intermediate level resolution of our model,
which does  not explicitly  resolves the body  surface (but  rather its
``hydrodynamic'' volume) this effective no-slip radius $R_{s}$ differs
from the  hydrodynamic radius  $R_H$ obtained from  its response  as a
monopole (Stokeslet) at long  distances. Differences in these long and
short range  radius estimations  are not large  and decrease  with the
particle size.  These are analyzed in Sec. \ref{sec:consistency} but before
that, we will show that mutual  mobilities, lubrication  and stresslet 
arising from higher multipole  contributions in the near-field 
are  consistent   with  the  same   effective  ``no-slip'' sphere radius.

Figure \ref{fig:vel} shows the  velocity profiles around the multiblob
icosahedron. These figures correspond to averages over many different
orientations of  the icosahedrons.  Comparisons are made  with the  flow past  a rigid
no-slip sphere with radius  $R_{s}$ (dashed lines), whose radial $v_r$
and tangential  $v_{\theta}$ components vary with the  distance to the
particle center $r$ as,
\begin{eqnarray}
  \label{vsphere}
  \frac{v_r}{u_0 \cos\theta}&=&1-\frac{3}{2}\frac{R_{s}}{r}
  \left[1-\frac{1}{3} \left(\frac{R_{s}}{r}\right)^2\right]\\
  \frac{v_\theta}{u_0\sin\theta} &=&1-\frac{3}{4}\frac{R_{s}}{r}\left[
    1+\frac{1}{3}\left(\frac{R_{s}}{r}\right)^2\right]
\end{eqnarray}
Here $\br$ is the radial vector with origin in the particle center and
$\theta$ is the angle between the particle velocity $\bU_0$ and $\br$.
Figure \ref{fig:vel} shows reasonably good fits to the sphere velocity
profiles of Eq.  (\ref{vsphere}).  The effective radius of the sphere
$R_s$  which  best  fits  the  multiblob velocity  profiles  at  close
distance  is indicated  in the  figure legend. The value of $R_s$ 
of the multiblob icosahedron model provides an estimation of its
``no-slip'' radius, compatible with a rigid sphere. The fitting procedure
involves adjusting several curves  with one only parameter ($R_s$)
whose fitting uncertainty is about  $0.05h$. 

We note that in these fits (see  Fig.  \ref{fig:vel})
we only consider the  velocity profile near the  particle, at
distances  of about  $r<2R_s$  where higher
moments (less sensitive  to  the effect  of  periodic images) become
important.   We observed that  the box  size $L$  does not affect the
estimation of $R_s$ more than the inherent
uncertainty of  the method.  As  a final  comment, we
find  remarkable the  extremely small  fluid leakage  observed  in the
largest particles  considered (see Fig.   \ref{fig:vel}), whose vertex
distances reach up to  $3.25\,h$ and correspond to non-overlapping blob
kernels (the blob kernel width is $3h$). 
This is probably due to the good behavior of Peskin's
kernels  and  opens  the possibility  of  studying  polidispersity
effects with the very same model,over a range of particle radius up to
about $3.5$ times larger than the blob.

\section{Mutual friction and lubrication forces}
\label{sec:fric}

The  mutual friction  induced by  hydrodynamic
coupling  between  two  particles is now analyzed.   More precisely,  by  pulling  two
particles  with  equal  by  opposite  sign forces  and  measuring  the
resulting linear  and angular particle velocities, we  obtain the pair
mobility  dependence  with   their  distance.   The  impact  parameter
(closest distance between linear trayectories) was  set to zero in a series
of  simulations  dedicated  to   lubrication  forces while, in another set
the   impact  parameter  was  chosen to be about  the hydrodynamic diameter.  In this second
set of simulations particles pass-by at close
distance which permit to measure the tangential component of  the mutual
mobility and the induced translation-rotation coupling.

\subsection{Normal, tangential friction and rotational coupling}

\begin{figure}
  \centering
  \includegraphics[width=0.7\textwidth]{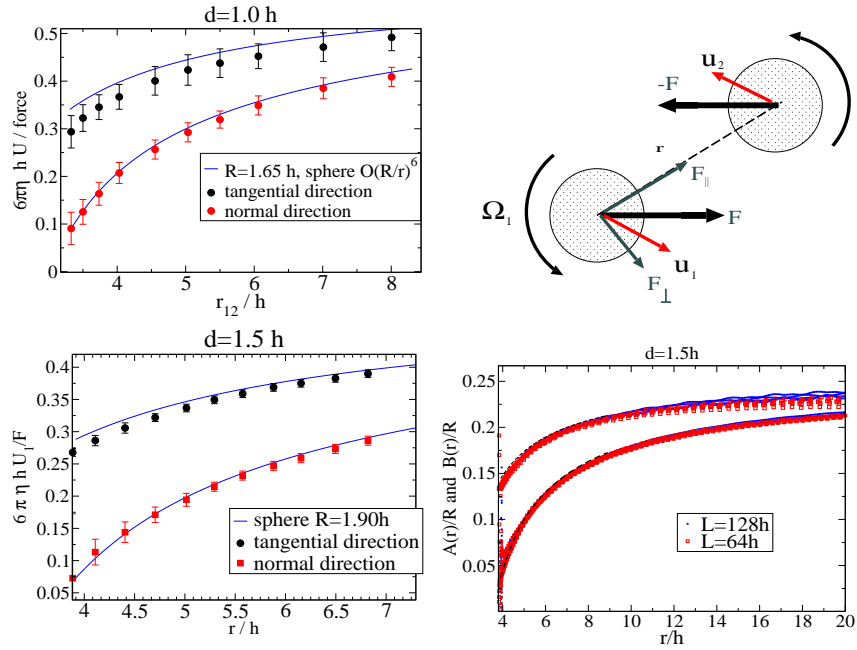}
  \caption{Mutual friction between two particles approaching at non-zero
    impact parameter as  specified in the sketch.  In  each pannel, the
    upper curves are obtained from the tangential mobilities and the lower
    ones from normal mobilities.  The tangential component of any vector
    (e.g. pulling  force) being  ${\bf F}_{\perp}\cdot \left({\bf  1} -
    {\bf e} {\bf e}\right)$ and the normal one ${\bf F}_{||}\cdot {\bf
      e}  {\bf e}$,  where ${\bf  e}={\br_{12}}/{r_{12}}$.   The dashed
    curves are fits to the multiple reflection expansion for
    two spheres (errors of order $O(R/r)^{6}$) in Eq. \ref{ab} using   values  of  sphere   effective  radius  $R$ 
    indicated in the legends.  The size of the box in the left panels is $L=64\,h$. In the right pannel (bottom) we
    compare results for $d=1.5\,h$ in two boxes $L=128\,h$ and $L=64\,h$.
    \label{fig:tang}
  }
\end{figure}

Here we present results for a  set  of simulations  where  two  particles
approach each other and pass by at close distance.  Both particles are
pulled  by  similar  forces  in opposite  directions  ${\bf  F_1}={\bf
  F}=-{\bf  F}_2$.   The  setup,  sketched in  Fig.   \ref{fig:tang},
corresponds  to  a finite  impact  parameter  of  about the  particle
diameter  (nearly touching spheres).  To compare the
results  of the  multiblob particles  with  rigid spheres  we use the
theoretical  result obtained  from the  method of  multiple reflections
(four reflections) valid up to $O(r^{-6})$ \cite{KimKarrila}
\begin{eqnarray}
  6\pi\eta R_{s} \bU_0^{(1)}&=& {\bf F} \cdot \left[ A(r) {\bf e_r e_r} + B(r) ({\bf 1} -{\bf e_r e_r})\right]\\
  \label{rot}
  6\pi\eta R_{s}^2 {\bs \Omega}^{(1)}&=& {\bf F}\times {\bf e_r} \frac{3}{4}\left(\frac{R_{s}}{r}\right)^2 
\end{eqnarray}
here   $\br$  is  the   vector  joining   the  two   particle  centers
$\br=\br_2-\br_1$  and  ${\bf e_r}=\br/r$.   The  angular velocity  of
sphere 1 is ${\bs \Omega}^{(1)}$  and equals $-{\bs \Omega}^{(2)}$.  The functions
$A(r)$ and $B(r)$ determine the amplitude of the normal and tangential frictions,   respectively   along  ${\bf   e_r}$   direction  and   the
perpendicular direction coplanar to $\bF$.  These functions are,
\begin{eqnarray}
  \label{ab}
  A(r)&=& 1 - \frac{3}{3}\left(\frac{R_{s}}{r}\right)   +\left(\frac{R_{s}}{r}\right)^3
-\frac{15}{4} \left(\frac{R_{s}}{r}\right)^4 \\
  B(r)&=& 1-\frac{3}{4} \frac{R_{s}}{r}-\frac{1}{2}\left(\frac{R_{s}}{r}\right)^3
\end{eqnarray}

Figure         \ref{fig:tang}        shows        $\widetilde{A}\equiv
6\pi\eta\bU_{0,||}/F_{||}$       and       $\widetilde{B}       \equiv
6\pi\eta\bU_{0,\perp}/F_{\perp}$ with $F_{||}=\bF \cdot {\bf e_r}$ and
$F_{\perp}=\bF - \bF \cdot {\bf  e_r} {\bf e_r}$.  A perfect agreement
with  the result  for sphere  of  radius $R_{s}$  would correspond  to
$\widetilde{A} \rightarrow  A(r)/R_{s}$ and $\widetilde{B} \rightarrow
B(r)/R_{s}$. As shown in the figure, 
for $d=h$, the value $R_{s}\simeq
1.65\,h$   provides  a   reasonable  good   fit  with   the  multiblob
icosahedron.  
The best correspondence for the $d=1.5\,h$ case is found
to be  for $R_{s}\simeq 1.90\,h$.  These values are within $3\%$ 
with those observed in the velocity profiles. Also  to check for finite box effect we
reproduced   some   of   the   cases   with   different   box   sizes.
Fig.  \ref{fig:tang} shows  a  comparison for  the mobility  functions
$A(r)$  and $B(r)$  in  the  $d=1.5\,h$ case  obtained  with boxes  of
$L=64\,h$ and  $128\,h$. The agreement  of both curves for  $r<10h$ is
perfect,  showing no  trace of finite box size effects. These effects
start to be visible for $r>10h$ which corresponds to about $r>5\,R_s$.

\begin{figure}
  \centering
  \includegraphics[width=0.5\textwidth]{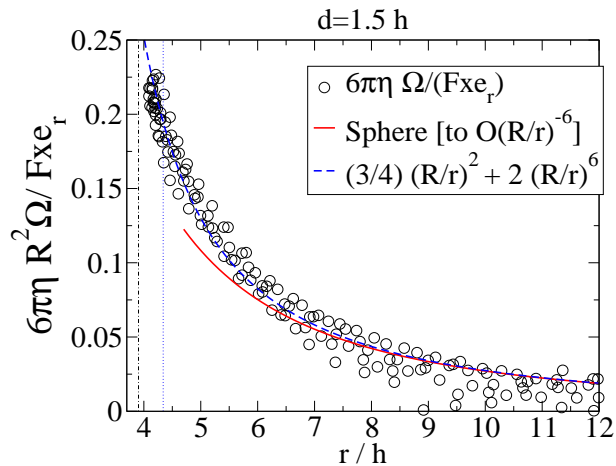}
  \caption{The  ratio  $6\pi\eta  R^2  {\bs \Omega}/|\bF\times  {\bf e_r}|$  where
    ${\bs \Omega}$  is  the  angular  velocity  of  one  sphere  induced  by
    hydrodynamic interaction with  another sphere passing close-by (see
    the  sketch of the  setup in  Fig.  \ref{fig:tang}).   Solid lines
    corresponds    to   the    sphere   result    (third   reflection)
    $(3/4)(R/r)^2$, which is valid up to $O(r^{-6})$.  The dashed line
    adds the next term in the expansion obtained by a fit to numerical
    results for which we have used the mobility effective radius $R=R_s=1.9\,h$.}
  \label{fig:rot}
\end{figure}

When  a particle  pass by near  by another a  torque is
induced  by  hydrodynamic  coupling.   In  terms  of  mobilities,  two
particles  of radius $R$ moving  nearby at distance $r$ 
by  a constant  force  $\bF$ experience  an
rotation   whose angular velocity (up   to  $O(R/r)^{6}$)   is   given   by
Eq.  (\ref{rot}).  To this  order in  the inverse  particles distance,
multiple reflections theory  \cite{KimKarrila} indicates that the ratio
between  the angular rotation  ${\bs \Omega}$  and the  induced torque
$\bF\times {\bf e_r}$ is independent on the particle radius (note that
$R_{s}^2$  appears  in   both  sides  of  Eq.  \ref{rot}).   Thus,
comparison with the multiple reflection solution to $O(r^{-6})$ serves
as a  check of the model  consistency at moderate  distances, but does
not bring  about any estimation of the  particle ``effective radius''.
Figure  \ref{fig:rot}  compares  both  signals  in  the  case  of  our
multiblob model for $d=1.5\,h$.  In these tests, the forces $\bF$ were
in  $x$-direction and the  angular velocity  was perpendicular  to the
${\bF}-\br$  plane.  The  other   two  components  of  ${\bs  \Omega}$
oscillate   around   zero.    For   $r>1.5R_s$  results   agree   with
Eq.  (\ref{rot})  while  at  shorter distances  the  angular  velocity
substantially   increase.   According   to  numerical   results   (see
Fig. \ref{fig:rot}) the  next term in the expansion  of Eq. (\ref{ab})
could be  $2(R_s/r)^{6}$ (with $R_s=1.9\,h$). In any case, the  increase in angular
velocity at close distances is qualitatively consistent with the weak
divergence of the angular velocity at short distances predicted by the
lubrication theory.

\subsection{Lubrication}
\label{sec:norm}

\begin{figure}
  \centering
  \includegraphics[width=0.5\textwidth]{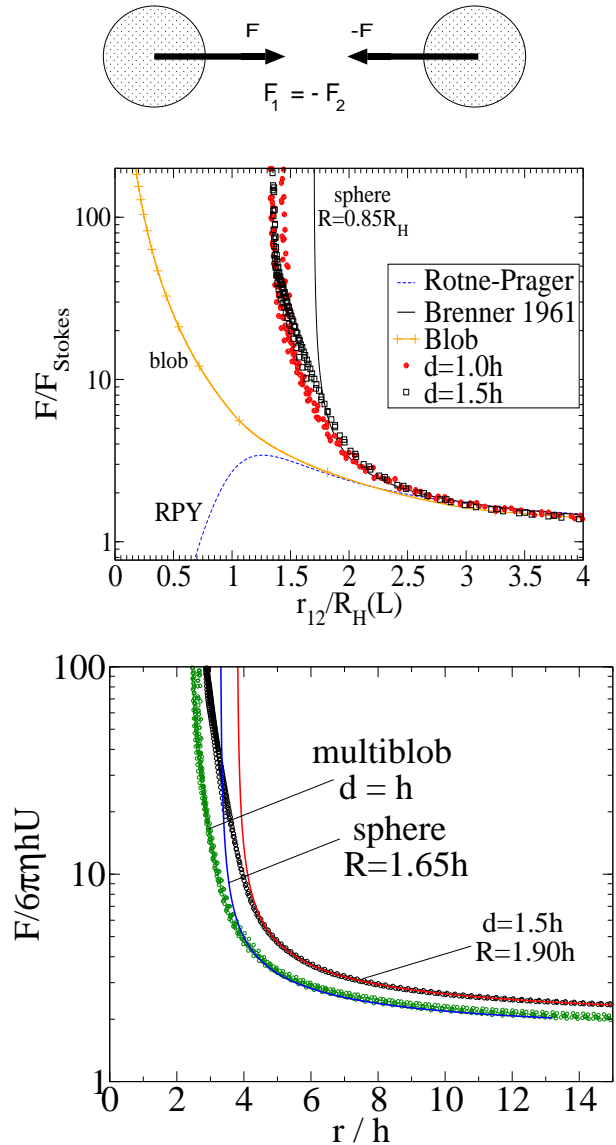}
  \caption{Drag force on one particle  due to the mutual friction with
    another  one   approaching  in  frontal   collision  (zero  impact
    parameter).Both particles being pulled  with the same force $F$ in
    opposite directions (see the  sketch).  Top pannel: friction force
    scaled  with   the  Stokes  force  against   the  scaled  distance
    $r/R_H(L)$.  At large  distances  $r>2.3R_H$ the  single blob  and
    multiblob scaled friction  coincide with the Rotne-Prager-Yamawaka
    analytic approximation.   For $r>1.85\,R_H$,  the multiblob  friction  agrees with
    Brenner's   \cite{Brenner1961}    analytic   solution   of   two
    approaching  spheres (we use an effective sphere radius  $0.85\,R_H(L) \simeq  R_s$).  
    Bottom pannel: the mutual
    multiblob friction scaled with  $6\pi\eta h u_0^{(1)}$ is compared
    with   Brenner's   analytic  result   for   two  rigid   spheres
    \cite{Brenner1961} with  effective no-slip radii  $R_s$, indicated
    in the legend. All multiblob simulations with $L=128\,h$ and $N=13$.
    \label{fig:lub}
  }
\end{figure}

Figure \ref{fig:lub} presents the results of two particles approaching
each  other with  zero impact  parameter.  As  stated, we  pulled both
particles with a constant force in opposite directions (i.e. the total
momentum  added to  the  flow  is zero)  and  calculate the  particles
velocities.   This  is  a   mobility  calculation,  although  in  Fig.
\ref{fig:lub} we present the results  in terms of friction (inverse of
the mobility), by  scaling the applied force $F$  with the velocity of
one of the  particles.  

The top pannel of Fig.\ref{fig:lub} presents the scaled friction force
$F/F_{Stokes}$  (with $F_{Stokes}=6\pi\eta R_H(L)  u_0^{(1)}$) against
the scaled distance $r/R_H(L)$.   This scaling permits to compare with
the single blob  behavior \cite{Balboa2012a,Balboa2014}, whose mobility is
compatible with Rotne-Prager-Yamawaka \cite{Rotne1969}.  Results for
multiblobs merge with the Rotne-Prager result for $r>2.5R_H$.  However
at shorter distances  a large increase of friction  is observed, which
seems to diverge at finite interparticle distance. This nice multiblob
``divergent  lubrication''  stems  out  from the  single  blob  mutual
friction, which (see Fig.  \ref{fig:lub}) is known also to ``diverge''
once two blob kernels fully overlap \cite{Balboa2012a}.  As expected, the
multiblob  divergence is  however not  fully compatible  with  a rigid
sphere.  This  is shown in Fig.  \ref{fig:lub}  by including Brenner's
analytic solution  for two approaching spheres \cite{Brenner1961} (see also 
Ref. \cite{Hansford1970}).   Using an effective
sphere  radius of  $0.85\,R_H(L)$ the  agreement with  a  rigid sphere
holds up  to distances of about $1.85\,R_H(L)$,  where friction forces
are about $6$ times larger than the Stokes drag.

It is noted that in using the term {\em divergent lubrication} for the
multiblob  we are abusing  the terminology  as, strictly  speaking, we
have not identified a coordinate (e.g. distance between closest blobs)
which  unambiguously determines  the  location of  the divergence  (see
Fig. \ref{fig:lub}).  In terms  of grid mesh units, the ``divergence''
of  multiblob  friction  was  observed  to occur  within  a  range  of
distances  $(2.6  \pm 0.2)h$  which  is  roughly  independent  on  the
icosahedron  size  $d$.   This  is  shown  in  the  bottom  panel  of
Fig. \ref{fig:lub} which presents the ratio $F/(6\pi\eta h U_0)$
(converging  to $R_H(L)/h$  at large  interparticle  separations $r$).
The  objective  of Fig.   \ref{fig:lub}  (bottom)  is  to compare  the
lubrication  of the  multiblobs with  that of  two spheres  having the
effective ``no-slip'' radius $R_s$  derived in previous sections (from
velocity  profiles  and  mobilities,  see  Figs.   \ref{fig:tang}  and
\ref{fig:vel}).   The  agreement   is  excellent  up  to  $r>2.2\,R_s$
indicating  the consistency  with near-field  hydrodynamics.  However,
due to  a ``softer''  hydrodynamic interaction, the  ``divergence'' of
multiblob  friction takes  place  at distances  somewhat shorter  than
$2R_s$.   In  the  next  section  we  show how  to  benefit  from  the
``natural'' lubrication of the model to recover the viscosity of dense
colloidal suspensions.

All cases  of Fig. \ref{fig:lub} corresponds to  the ``filled'' $N=13$
multiblob.   However,  it is  interesting  to  note that  calculations
performed  with the  $N=12$  icosahedron ``shell''  reported the  same
outcome for  the mutual friction,  meaning that the large  increase in
lubrication between  two multiblobs  is essentially determined  by the
external blobs and not by the inner blob.

\section{Stresslet and Viscosity}
\label{sec:stresslet}

This  section  analyzes  an  extremely  important  property  of  rigid
particles, which is  their ability of exerting a  finite stress in the
fluid.  In particular,  rigid  particles cannot  deform  and impose  a
constraint on the local  fluid velocity gradients, which vanish inside
the  particle domain.   This  requires  a work  done  by the  particle
cohesion forces,  which appears as an  extra virial term  in the fluid
stress   tensor    coming   from   the    fluid-particle   interaction
\cite{KimKarrila}.     This    virial    contribution,    derived    in
Eq. (\ref{presp})  for the present  model, increases the  fluid stress
and also the  effective viscosity of the colloidal  solution, which is
nothing  but  the ratio  between  the  stress  and the  fluid  overall
deformation rate.   For the reasons explained  in Sec. \ref{stresslet}
the contribution  of particle shape fluctuations in  the viscosity are
vanishingly small and they behaves like rigid spheres.

The viscosity  of a colloidal solution  is a long standing  problem which
have been  studied by many  authors \cite{PhysToday_WagnerBrady}.  The  problem is
usually  posed in the  low Reynolds  limit $\mathrm{Re}=\rho_f  V R/\eta
<<1$, and this leaves at  least two controlling parameters: the volume
fraction of  the colloids $\phi$ and the  Peclet number $\mathrm{Pe}=V
R/D$, with $D=\kt/(6\pi\eta R)$ the colloid diffusion coefficient.
The fluid environmental  velocity $V$ depends on the type of flow considered,
here, we have used a protocol to measure the viscous contribution of the particles 
in the limits $\mbox{Pe}=0$ and Schmidt number $\mbox{Sc}=\eta / (\rho D)  \rightarrow \infty$.

The viscosity was measured using a standard non-equilibrium procedure;
i.e., by exerting a periodic density force to the fluid ($\bs{f}(\br,t)=f_0 \sin(k_j y) \bs{\hat{x}}$)
and measuring the response of the velocity field in the $x$ direction. The viscosity
is the ratio
\eqn
\eta = \fr{f_0}{k_j^2 v_x}, 
\eqnend
where $v_x$ is the amplitude of the sinusoidal $x-$velocity profile 
formed along the $y$ direction. 
The wavenumber is $k_j=2j\pi/L$ with  $j$ a natural number so that the 
wavelength $\lambda=L/j$ fits in the periodic box of side $L$. 

In  calculating the stress contribution from
the particles one needs to sample all possible particle configurations
in the fluid which in  a straightforward simulation would imply letting
the   particles   diffuse   over   the  system   and   explore   their
configurational and positional phase  space.  Indeed, due to the large
time separation  between the spring dynamics and the particle diffusion dynamics, this
straight procedure is far from efficient, specially in the limit $\mbox{Pe} \ll 1$.
Instead, we decided to adopt a sort  of mixed  Monte Carlo approach. 

Initially,  random  configuration  on non-overlapping  particles  were
generated for a given concentration  $\phi$ and box size $L$ using the
Monte Carlo method.   No shear was introduced during  this part of the
process and  therefore the  Peclet number was  strictly zero.  In this
preparatory step, the particle  interaction potential was chosen to be
the hard-sphere potential so as to avoid any influence of the shape of
continuous (soft) repulsive cores on the calculation. The radius of the
hard-sphere interaction is  noted as $R_{\tex{HS}}$ and it  is indeed a free
parameter, such as any other set of possible interparticle interaction
parameters  (depletion, electrostatic,  etc.) would  be.  We  chose a
collection   of  independent   configurations   compatible  with   the
hard-sphere  potential  and the  thermodynamic  state,  and for  every
configuration we ran a short simulation in the presence of shear so as
to  measure the  effect  of the  extra  stress on  the  system on  the
amplitude of  the velocity profile.   To avoid a long  relaxation time
coming  from  the   fluid  inertia,  for  this  second   set  of  
{\em hydrodynamic}   simulations  we   used   the  
Fluctuating Immersed Boundary method for Brownian hydrodynamics that we  have recently  derived
 \cite{Stokeslimit}.  
This algorithm works in the  limit $\mbox{Sc}\rightarrow \infty$  in which
the relaxation of the fluid  momentum is instantaneous. However, we let
the initial configuration  relax over  a small  set of  iterations to
allow the  particles springs to adapt to  the flow strain:  the system is
relaxed over $4000$ time steps  and the velocity profile averaged over
$100$ time steps ($\dt=10^{-3}$).

This method proved to be quite efficient for the present purposes.
In particular it permits to surpass the bottleneck of the long
diffusive sampling times, which are dominant at $\mathrm{Pe}\rightarrow 0$
(the situation is not that critical as  $\mathrm{Pe} >1$).
In this study particles do not interact via potential forces, however, 
the same idea could be in principle extended to interacting particles, provided
the particle configurations are sampled from a Monte Carlo code
and sequentially fed to the hydrodynamic solver. This is so because, physically,
the positional and configurational probability distributions
does not depends on the hydrodynamics in the regime of low Peclet numbers.

Results for $\eta$ presented below are robust against changes in the wavelengths $\lambda$ of the external force, 
which we varied to check that  the measured viscosity 
corresponds  to the  $\lambda\rightarrow \infty$  (macroscopic) limit (see Fig. \ref{fig:vis}).
The  viscosity  of  molecular  liquids  is  known  to  depend  on  the 
wavenumber and $\eta(k)\rightarrow  \eta(0)$ for $kR \simeq 1$ \cite{Hansen_Simple_Liquids}. 
Although  this effect in colloidal hydrodynamics is potentially interesting by itself, 
here $\lambda$  was chosen much larger  than the colloidal size to 
avoid  observing any dependence of $\eta$ with the perturbative flow wavelength. 
Also, we varied $L$ for fixed $\lambda$ to test any possible finite size effect due to the box periodicity. The results for $\eta(\phi)$ 
in Fig.  \ref{fig:vis}), are quite satisfactory in these respects, showing that finite  
size effects are absent (either in $\lambda$ or $L$).  

\subsubsection{Dilute and semidilute regime}

Values of  the colloidal solution viscosity  $\eta(\phi)$ obtained for
the  $d=h$  icosahedron  are  reported  in  Fig.  \ref{fig:vis}.   The
viscosity increases with the particle volume fraction $\phi= (4/3) \pi
R_{\tex{HS}}^3 n  L^{-3}$ with  $n$ the number  of particles in  the system.
Using  $R_{\tex{HS}}=1.62\,h$ we  recover  an excellent  agreement with  the
classical Einstein relation
\begin{equation}
  \eta=\eta_0 \left(1+\frac{5}{2} \phi\right),
\end{equation}
which  is  valid  at very  low  dilutions,  as  can  be also  seen  in
Fig. \ref{fig:vis}. At higher concentration of colloids, the viscosity
has  also   contributions  from  the   hydrodynamic  coupling  between
particles \cite{KimKarrila,Batchelor} which induce a quadratic term in the
virial expansion of $\eta$. The result, originally derived by Batchelor \cite{Batchelor} is 
\begin{equation} 
  \eta =\eta_0 \left(1+\frac{5}{2} \phi + 6.2 \phi^2 + ...\right). 
\end{equation} 
The quadratic  regime in the viscosity solutions  of colloidal spheres
is  also shown in  Fig. \ref{fig:vis}. Quite importantly, the theoretical
sphere trend is recovered for $R_{\tex{HS}}=1.62\,h$, which is in excellent  agreement 
with values of $R_s$ obtained from velocity profiles and mutual friction
$R_s=(1.63 \pm 0.03)\,h$.

\subsubsection{Dense regime}
\begin{figure}
  \centering
  \includegraphics[width=0.5\textwidth]{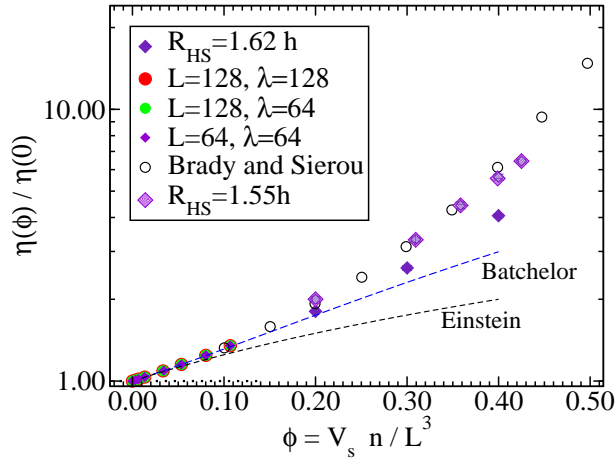}
  \caption{\label{fig:vis}  Intrinsic  viscosity  of a  dispersion  of
    multiblob   particles  at   increasing  volume   fraction  $\phi$.
    Simulations  correspond to  $d=h$ in  periodic boxes  of different
    length, imposing a  sinusoidal $\sin(2\pi y/\lambda)$ shear stress
    acting over  $x$ direction.  The  best fit to the  Einstein linear
    regime  $\eta(\phi)=\eta_0\left(1+2.5\phi\right)$  at  low  volume
    fraction   $\phi$   provides   the   effective   particle   volume
    $\mathcal{V}_s  =(4\pi/3)\,R_{s}^3$  compatible  with  the  sphere
    behavior.   The   effective    particle   radius   resulting   in
    $R_{s}=1.62\,h$. 
    At larger  $\phi<0.15$  we   find  excellent
    agreement  with the  Batchelor regime.  For denser  suspensions we
    compare the results obtained  for hard-spheres of radii $R_{\tex{HS}}$,
    with  the fully  resolved simulations  result of  Brady and
    Sierou \cite{BradySierou2001}.  Multiblobs  with $R_{\tex{HS}}=1.55h$ are in
    very good agreement up to $\phi \simeq 0.45$.}
\end{figure}

As shown  in Fig.\ref{fig:vis}, for $\phi>0.2$  the viscosity obtained
for  multiblobs  with hard-spheres  of  radii $R_{\tex{HS}}=R_s=1.62\,h$  is
somewhat  below  the  sphere   trend  reported  by  Brady  and  Sierou
\cite{BradySierou2001}. This was to be expected because although multibody
hydrodynamic  interactions  are taken  into  account, the  lubrication
curve  for  our  multiblob   model  peaks  off  (or  ``diverges'')  at
interparticle distances  smaller than  $2\,R_s$.  In other  words, for
$R_{\tex{HS}}=1.62\,h$ the large lubrication regime of the model is screened
by the  particles excluded  volume.  However, to  deal with  the dense
regime one  can slightly modify  the hard-sphere interaction so  as to
allow  the  model unfold  the  required  amount  of lubrication.   In
particular,  to   resolve  the  large  lubrication   regime  we  tried
multiblobs  with  hard-sphere   radius  $R_{\tex{HS}}=1.55\,h$  (instead  of
$1.62\,h$).  Results,  shown in Fig.   \ref{fig:vis} are in  very good
agreement with rigid spheres up to $\phi \simeq 0.45$. We believe this
is an outstanding result as  it naturally stems from the fluid solver,
and    does   not    requires   ad    hoc    lubrication   corrections
\cite{Nguyen2002,Jens2013}.   At this  highly  concentrated solutions,
the  hydrodynamic  interaction  is  essentially short-ranged,  so  the
mismatch between  the potential  interaction radius $1.55\,h$  and the
hydrodynamic radius  of one isolated particle is  not really important
for the physics  involved.  We have not tried  to further increase the
particle volume  fraction towards $0.5$  (in fact, the  preparation of
the initial configuration becomes really hard); however to resolve the
lubrication  of this  extremely dense  regime we  could  still further
reduce the excluded volume radius to about $R_{\tex{HS}}\simeq 1.50\,h$.

\section{Consistency in the particle hydrodynamic size}
\label{sec:consistency}

Contrary      to       other      models      in       this      field
\cite{Uhlmann2005,Peskin2002,Feng2005,Molina-Yamamoto2013,Jens2013},
our multiblob design  is not based on resolving  the particle surface,
but rather its core, and this necessarily implies some uncertainty its
hydrodynamic  size.   This section  compares  the  results for  filled
icosahedrons multiblobs  of different sizes  (vertex distances between
$d=h$ and $d=3.25h$) and fixed number of blobs ($N=13$).

\begin{table}[h]
  \caption{Radii of the multiblob  icosahedrons of vertex distance $d$
    obtained from  different routes. Second column;  radius of inertia
    $R_I$ evaluated by fitting the  moment of inertia of the multiblob
    icosahedron     to     a     sphere     $R_I     =     \left((5/2)
    \mIt/M\right)^{1/2}$.   The  mass   $M$  is   obtained   from  Eq.
    \ref{mass2}.  Errors  bars in  $R_I$ are around  $0.01\,h$.  Third
    column:  hydrodynamic radius  $R_H(L\rightarrow  \infty)$ obtained
    from the  Stokeslet response.  Fourth column: rotlet  radius $R_r$
    from rotational diffusion  (see Sec. \ref{sec:rot}). Fifth column:
    the {\em no-slip}  radius obtained from best fits of the sphere colloids
    near-flow perturbation,  mutual  mobility,  lubrication  and Stresslet. All quantities in unit of the Eulerian grid
    mesh $h$.}
  \begin{tabular}{|c|c|c|c|c|}
    \hline
    $d$ & $R_I$ & $R_H$ & $R_r$ &  $R_s$        \\
    \hline 
    \hline                        
    0 (blob)   &   -      &   0.91 &    -  &  -    \\
    1.0 &  1.61         &   1.76 &  1.50 & $1.62 \pm 0.03$  \\ 
    1.5 &  1.85         &   1.99 &  1.90 & $1.87 \pm 0.03$  \\ 
    2.0 &  2.18         &   2.35 &  2.36 & 2.30  \\ 
    2.5 &  2.67         &   2.73 &  2.63 & 2.70  \\ 
    3.0 &  3.30         &   3.08 &  3.30 & $3.00 \pm 0.05$  \\ 
    \hline                               
  \end{tabular}
  \label{tab:rad}
\end{table}

To begin with, the radius  of inertia of the particle $R_I$ calculated
in  Sec. \ref{sec:rot}, is just  a property of the particle shape (more
properly,  of  the distribution  of  the  forming  blobs) and  of  the
interpolators used (here 3 pt  Peskin kernels). $R_I$ can be sought as
a reference particle size, which should ideally be consistent with the
disturbances created  by a multiblob  in the fluid. 

The hydrodynamic response of  the multiblob essentially depends on the
relative importance  of the multipole terms of  the perturbative flow.
The  multiblob monopole  is relevant  at long  distances and  was made
compatible  with the  Stokeslet of  a rigid  no-slip sphere  of radius
$R_H$. Following the standard procedure \cite{Duenweg2009}, we call $R_H$ the
multiblob hydrodynamic radius,  which depends on the box  size $L$ due
to well established finite size effect in periodic space.  As shown in
Table  \ref{tab:rad}, we  found  that for  $d<  3h$, the  hydrodynamic
radius $R_H(L\rightarrow\infty)$ is slightly larger than $R_I$.

Hydrodynamic properties, such as  the the fluid velocity profiles past
a fixed multiblob  and the mutual friction are  much less sensitive to
finite  size  effects,  reflecting  the relevance  of  the  near-field
response dictated  by higher order terms of  the multipolar expansion.
In all these tests we found that the multiblob behavior is consistent
with an effective  sphere of radius $R_s$.  Lacking  a better name, we
called $R_s$  the effective  {\em no-slip} radius  and found  that the
validity  of this ``short-range''  regime extends  up to  distances of
about  $r\sim  5\,R_s$  away  from  the particle  center  (see  Figure
\ref{fig:lub}).  Consistently  with this observation,  the stresslet in
the dilute regime  was also found to be  compatible with rigid no-slip
spheres of radius $R_s$.  The same applies for the rotational friction
$R_r$ which we found in relative good agreement with $R_s$.

\begin{figure}
  \includegraphics[width=0.4\textwidth]{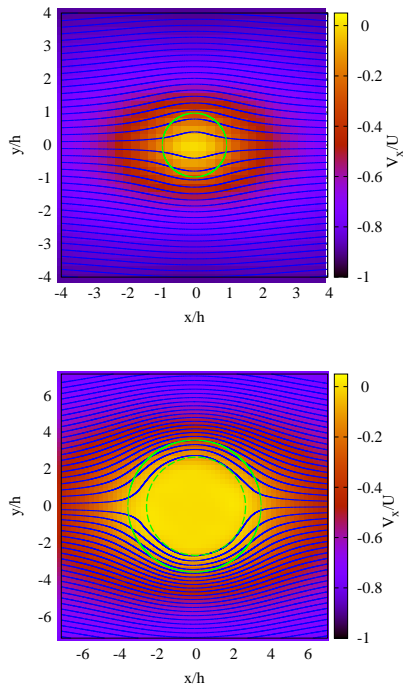}\\
  \caption{\label{fig:stream}  The streamlines  around a  fixed single
    blob (top)  facing a plug flow,  compared with those  created by a
    filled icosahedron formed by $N=13$  blobs (one in its center) and
    inter-vertex distance  $d=3h$.  The Reynolds number  is very small
    so the flow is symmetric  in the fore and aft-directions.  Colours
    indicate isovalues  of the flow  velocity in the direction  of the
    particle motion ($x$). The solid green circle around the multiblob
    particle  indicates  the   location  of  its  hydrodynamic  radius
    $R_H(L)$ (with  $L=64h$ in the  figure) and the dashed  circle the
    location where the  fluid relative velocity approximately vanishes
    ``(effective no-slip radius, $R_S$)''.}
\end{figure}

Figure   \ref{fig:stream}  illustrates  the  flow  streamlines  past  a
multiblob particle, and compares it with the single blob case. 
The hydrodynamic radius $R_H$ and the  no-slip  radius  $R_s$
have been indicated with lines. 
The iso-values of the fluid  velocity in $x$ direction, 
clearly indicates that at  $r\simeq  R_s$  the
gradient of the stream-function in normal direction vanishes, meaning
zero fluid-particle  relative velocity.  This outlines
the particle ``core'', as shown in Fig. \ref{fig:stream}. By contrast the
blob model lacks a core domain in proper sense.
In Fig. \ref{fig:stream} we have chosen a relative
large particle $d=3h$ in a small box $L=64h$ to illustrate the case
with the largest difference in $R_H(L) \simeq 3.5h$ and $R_s\simeq 3.0h$. This difference
($0.5h$ in the example) decreases with $L$ as $R_H(\infty)<R_H(L)$) but {\em also} 
with the particle size $a$. To understand the overall trend of the
model's radii it is illustrative to derive the so-called
Fax\'en radius of the multiblob kernel, $a_{F,m}$. We get (see Appendix),
\begin{equation}  
  \label{jm2}
  a_{F,m}^2= \Jm \left[ \left(\br -\bq_0\right)^2\right] = a_F^2 + a^2.
\end{equation}
were we recall that $a=0.9511\,d$ is the embedding sphere radius of the icosahedron
and, similarly,  $a_F^2$ is the Fax\'en square radius of the blob kernel (i.e. its second  moment) \cite{Balboa2014,Stokeslimit} given by,
\begin{equation}
  \label{jblob2}
  a_F^2  \equiv 3 \int \delta(\br) r_{\alpha}^2 dr^3 = \J\left[\br^2\right].
\end{equation}
Its value (for the 3pt kernel) is $a_F \simeq (0.95 \pm 0.05)\,h$, quite  close to the blob Stokeslet radius $R_H=0.91\,h$
\cite{Stokeslimit,Balboa2014}.

\begin{figure}
  \includegraphics[width=0.6\textwidth]{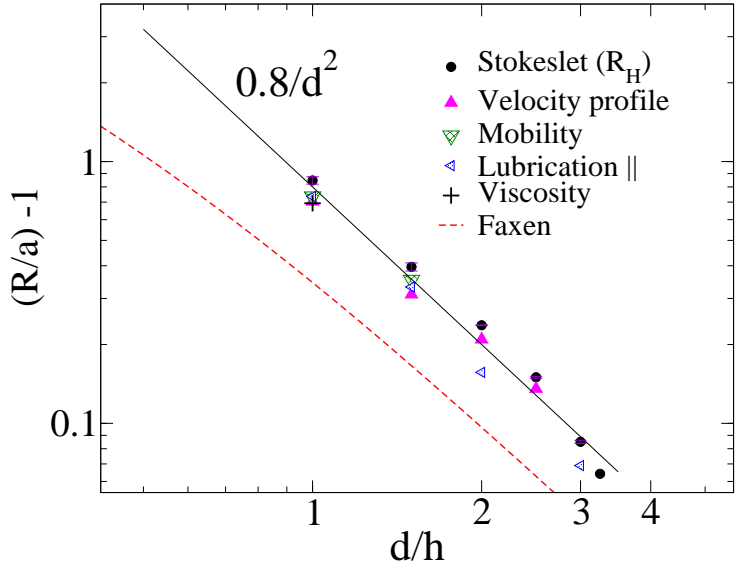}
  \caption{Relative difference  between the effective  particle radius
    estimated by comparison with  the sphere behavior under different
    physical mechanisms $R$ and  the geometric radius of the multiblob
    icosahedron particle.
    \label{fig:radius}
  }
\end{figure}

The  relation  (\ref{jm2})  indicates  that  the  relative  difference
$(a_{F,m}-a)/a$ decreases with the  geometric radius $a$, scaling like
$a^{-2}$. Such scaling corresponds to  the ratio between the volume of
the blob kernel  shell $4\pi a^2 a_{F}$ and  the total particle volume
$4(\pi/3)  a^3$.  Thus the deviations found
in the effective radii decrease quadratically
with  the multiblob size $a$, as highlights Fig.   \ref{fig:radius}.
The result  of Eq. (\ref{jm2})  together with Figs.  \ref{fig:rh} and
\ref{fig:radius} suggest ways to reduce the uncertainty in size:
for instance using higher order kernels for the blobs
in  shell of the particle to reduce $a_F$ and sharpen
the body surface to get $a\simeq  a_{F,m}\simeq R_H\simeq R_s$.
This however, might spoil the good lubrication properties of this model,
although it is something to be explored in future works.

We conclude  this section  with some general  comments on what  is the
best choice for the value of $d$.  The first issue to indicate is that
the size of the simulation box has to be $L\propto R_H$ while the
computational cost (fluid cells)  scales at least like $L^3$.  
Thus small particles  (i.e.  small values of
$d$)  reduce  simulation  costs. To give one example,  the
smallest  particles used in Refs.  \cite{Nguyen2002,Molina-Yamamoto2013,Breugem}  
are about  $R=8\,h$, while using $d=h$  ($R_H   \simeq  1.6\,h$) we reduce 
the number of  fluid cells embedding the colloids 
in about $(8/1.6)^3=125$ times. Another  benefit of
small multiblobs is that  their lubrication is better resolved due
to  the shorter difference between  their hydrodynamic  size  and the
interparticle  distance  where the mutual  friction  diverges (which  as
stated, was observed to be  roughly independent on $d$).  Values of $d$
between $h$ and $1.5\,h$ provide similar good outcomes in this respect (see
Fig.\ref{fig:lub}).   Unfortunately,   for  $d<h$  substantial  kernel
overlap  increases  the grid  dependence  of  the particle  properties
(mass, $R_H$, $R_I$).

Finally, an important issue to be considered is the consistency in the
particle  size,  in response  to  different hydrodynamic  interactions
(translation, rotation, mutual friction, etc.).  In this respect, the
main result of the present study is summarized in Table \ref{tab:rad}.
It is noted that $R_I$ is  only relevant for inertial effects related to rotation, 
so at small particle Reynolds number, and  in the dilute or moderately dense
colloidal solutions, the range $1.5\leq  d/h \leq 2.5$ provides a save
choice with differences in radii of less than about $5\%$.  The value
$d=2.5h$ provides a particularly consistent set of radii (also with $R_I$) as  can be seen
in Table \ref{tab:rad}.

\section{Conclusions}
The  present multiblob  model  targets an  intermediate resolution  of
colloidal particles, using a few  overlapping blob kernels (here 12 or
13) to  construct relatively small particles. Using  an icosahedron as
template  we get  particles radius  between $1.6h$  and  $3.25h$, with
hydrodynamic  response  compatible with  spherical  colloids (with  or
without  inertia), in  terms  of translation  and  rotation and  their
couplings, flow perturbation,  mutual friction and lubrication forces.
The   hydrodynamic   radius  determined   from   the  Stokeslet   [see
  Eq. (\ref{rhinf})] is known to depend  on the system size $L$ due to
periodic   images.   However,   hydrodynamic  interactions   at  short
distances  are  much  less  sensitive  to  finite  size  effects  (see
e.g. Fig. \ref{fig:tang}), and the effective particle sizes related to
rotation,  near-flow perturbative flow  and short  distance mobilities
agree within about $5\%$ for any $L$.

In the  method presented in  this paper the  hydrodynamic interactions
emerge  naturally  from  the   fluid  solver  and  the  fluid-particle
coupling. Notably, our schemes do not need to include ad hoc pair-wise
friction patches to model lubrication. The strength of the lubrication
force between  two multiblobs can be calibrated  by slightly modifying
their hard-core (or hard-sphere) 
radius $R_{HC}$ (more precisely  the ratio $R_{HC}/h$
which   is  a free non-hydrodynamic parameter).    In
particular, we have  explored the case of colloidal  spheres and were
able  to  reproduce the  increase  in  viscosity  up to  large  volume
fractions $\phi\leq 0.45$.  In future  work this idea will be extended
to  lubrication in  colloids  with complex  shapes, where  theoretical
lubrication relations  are difficult to implement or  even absent.  On
the  other  hand,  the  method   allows  for  an  easy  and  efficient
implementation in systems with periodic boundary conditions, for which
the  incompressibility constraint  can  be applied  with Fast  Fourier
Transform methods \cite{Balboa2014}.  In the viscosity measurements of
Sec.   \ref{sec:stresslet}  our implementation  of  these schemes  for
Graphical  Processing Units \cite{Usabiaga}  is able  to run  at about
$100$ time  steps per second for  a system of $1000$  icosahedron at a
volume  fraction  of $\phi  \approx  0.3$  in  one standard  GPU  card
(1.4GB);  this is about  $40$ times  faster than  reported simulations
using  Accelerated  Stokesian Dynamics  in  a  $16$  CPU core  cluster
\cite{Yeo2010}.  This  increased efficiency is important  to allow for
larger  temporal  windows  of  observation, necessary  to  study  some
non-equilibrium     processes    such    as     clustering    gelation
\cite{Candia2009,Whitmer2011,Cao2012}.

\section*{Acknowledgments}
We  thanks Aleks  Donev for  a critic  reading of  the  manuscript and
suggestions.   We  acknowledge  funding  from the  Spanish  government
FIS2010-22047-C05,   from   the   Comunidad  de   Madrid   MODELICO-CM
(S2009/ESP-1691), Acknowledgment is made to the Donors of the American
Chemical Society  Petroleum Research Fund for partial  support of this
research.

\appendix
\section{Multiblob Fax\'en radius}
We showed in previous works for single blob particles 
\cite{Balboa2014,Balboa_tesis} that there is a close
relation between the second moment of the kernel and its Fax\'en radius.
We shall show that in the Fax\'en radius of the multiblob 
is given by the second moment of the average kernel $\Jm$.
For a general body shape it consists on a dyadic, which can be directly
calculated from the definitions in Eqs. (\ref{ju}), (\ref{delta}) and (\ref{jm}),
and using the zeroth and linear consistency of the 3pt Peskin's (blob) kernel,
\begin{equation}
  \label{jm2a}
  \Jm\left[ \left(\br -\bq_0\right)^{\alpha}\left(\br -\bq_0\right)^{\beta}\right]
  = \frac{a_F^{2}}{3} \delta_{\alpha,\beta} + \frac{1}{N}\sum_i s_{i}^{\alpha} s_{i}^{\beta}
\end{equation}
Note that the second term in (\ref{jm2a}) is just the gyration tensor of the structure.
For a platonic solid, such as our icosahedron, $\sum_i s_{i}^{\alpha} s_{i}^{\beta} = (a^2/3)\, \delta_{\alpha,\beta}$
where $a$ is the radius of the sphere passing through the vertexes
(for our icosahedron $a=\sqrt{(10+2\sqrt{5})/4} \; d = 0.9511\,d$). In this case, 
the second moment tensor  reduces to a scalar times the identity matrix, its trace being
\begin{equation}  
  \label{jm2AP}
  \Jm \left[ \left(\br -\bq_0\right)^2\right] = a_F^2 + a^2.
\end{equation}
where  $a_F^2$ is the second  moment of  the  single blob defined in Eq. (\ref{jblob2}).
The result of Eq. (\ref{jm2AP}) is interesting because it indicates that the 
average kernel size (given by its second moment) and the geometric particle size $a$
converge as the particle size is increased, with {\em fixed number of blobs}.

The Fax\'en law for particle translation \cite{Dhont_book1996,KimKarrila}
can be used to relates the trajectory of a force-free, inertialess particle in any flow field. 
To relate the Fax\'en law with the kernel properties we proceed like in Refs. \cite{Stokeslimit,Balboa2014,Balboa_sound}:
we Taylor expand the flow field $\bv(\br)$ around the particle center $\bq_0$ 
and interpolate it on the particle site to obtain $\bU_0=\Jm[\bv(r)]$. One gets,
\begin{equation}
  \label{vfax}
  v^{\alpha}(\br) = v^{\alpha}(\bq_0) + \partial_{\beta} v^{\alpha}(\bq_0)\, (r-q_0)^{\beta} +
  \frac{1}{2} \partial_{\beta} \partial_{\gamma} \bv^{\alpha}(\bq_0) \, (r-q_0)^{\beta} (r-q_0)^{\gamma} + ...
\end{equation}
Applying the interpolation $\Jm$ and using the result of Eqs. (\ref{jm2a}) and (\ref{jm2AP}),
\begin{equation}
  \label{ufaxen}
  \bU_0=\Jm[\bv(\br)]  =  \bv(\bq_0) + \frac{a_F^2+a^2}{6} \nabla^2 \bv(\bq_0) +...  
\end{equation}
This result exactly corresponds to the Fax\'en law \cite{Dhont_book1996} 
on a force-free, inertialess  particle    of   radius  $a_{F,m}= (a_F^2+a^2)^{1/2}$ moving in the  fluid field  $\bv(\br)$. 
Any curvature in the velocity field (here proportional to the local total pressure gradient $\nabla^2 \bv = (1/\eta) \nabla p$)
induces a departure from the local fluid velocity proportional to the size of the particle. This observation, initially made for
the single blob case \cite{Balboa2014,Stokeslimit} led us to relate the second moment of the kernel to its  
``Fax\'en  radius''. We also observed that small kernels with non-zero second moment
are paradoxically better suited to represent small particles of finite size
(for the 3pt kernel $a_F \simeq (0.95 \pm 0.05)\,h$, which is close to its Stokeslet radius $R_H=0.91\,h$).

In  passing  we   note  that  for   non-isotropic  particles,
Eq.  (\ref{jm2a})  into  (\ref{vfax})  shows that  departures  of  the
particle  trajectory   from  the  unperturbed   fluid  trajectory  are
proportional  to  the  contraction   of  the  triadic  $\nabla  \nabla
\bv(\bq_0)$  with   the  gyration  tensor  of   the  structure  $(1/N)
\sum_i{\bf s}_i {\bf s}_i$.

\bibliography{icosahedron}

\begin{thebibliography}{51}
\expandafter\ifx\csname natexlab\endcsname\relax\def\natexlab#1{#1}\fi
\expandafter\ifx\csname bibnamefont\endcsname\relax
  \def\bibnamefont#1{#1}\fi
\expandafter\ifx\csname bibfnamefont\endcsname\relax
  \def\bibfnamefont#1{#1}\fi
\expandafter\ifx\csname citenamefont\endcsname\relax
  \def\citenamefont#1{#1}\fi
\expandafter\ifx\csname url\endcsname\relax
  \def\url#1{\texttt{#1}}\fi
\expandafter\ifx\csname urlprefix\endcsname\relax\def\urlprefix{URL }\fi
\providecommand{\bibinfo}[2]{#2}
\providecommand{\eprint}[2][]{\url{#2}}

\bibitem[{\citenamefont{Kim and Karrila}(1991)}]{KimKarrila}
\bibinfo{author}{\bibfnamefont{S.}~\bibnamefont{Kim}} \bibnamefont{and}
  \bibinfo{author}{\bibfnamefont{S.}~\bibnamefont{Karrila}},
  \emph{\bibinfo{title}{Microhydrodynaics:Principles and Selected
  Applications}} (\bibinfo{publisher}{Butterworth Heinemann},
  \bibinfo{address}{Boston}, \bibinfo{year}{1991}).

\bibitem[{\citenamefont{Lowen}(2012)}]{Lowen2012}
\bibinfo{author}{\bibfnamefont{H.}~\bibnamefont{Lowen}},
  \bibinfo{journal}{Journal of Physics: Condensed Matter}
  \textbf{\bibinfo{volume}{24}} (\bibinfo{year}{2012}).

\bibitem[{\citenamefont{Eaton}(2009)}]{Eaton2009}
\bibinfo{author}{\bibfnamefont{J.~K.} \bibnamefont{Eaton}},
  \bibinfo{journal}{International Journal of Multiphase Flow}
  \textbf{\bibinfo{volume}{35}}, \bibinfo{pages}{792} (\bibinfo{year}{2009}),
  ISSN \bibinfo{issn}{03019322},
  \urlprefix\url{http://linkinghub.elsevier.com/retrieve/pii/S0301932209000330%
}.

\bibitem[{\citenamefont{Cate et~al.}(2004)\citenamefont{Cate, Derksen, Portela,
  and den Akker}}]{Cate2004}
\bibinfo{author}{\bibfnamefont{A.~T.} \bibnamefont{Cate}},
  \bibinfo{author}{\bibfnamefont{J.~J.} \bibnamefont{Derksen}},
  \bibinfo{author}{\bibfnamefont{L.~M.} \bibnamefont{Portela}},
  \bibnamefont{and} \bibinfo{author}{\bibfnamefont{H.~E. A.~V.}
  \bibnamefont{den Akker}}, \bibinfo{journal}{J. Fluid Mech.}
  \textbf{\bibinfo{volume}{519}}, \bibinfo{pages}{233} (\bibinfo{year}{2004}).

\bibitem[{\citenamefont{Dhont}(1996)}]{Dhont_book1996}
\bibinfo{author}{\bibfnamefont{J.~K.~G.} \bibnamefont{Dhont}},
  \emph{\bibinfo{title}{{An Introduction to Dynamics of Colloids}}},
  vol.~\bibinfo{volume}{2} of \emph{\bibinfo{series}{Studies in Interface
  Science}} (\bibinfo{publisher}{Elsevier}, \bibinfo{year}{1996}),
  \bibinfo{edition}{studies in} ed., ISBN \bibinfo{isbn}{9780444820099}.

\bibitem[{\citenamefont{Happel and Brenner}(1983)}]{Happel-Brenner-book}
\bibinfo{author}{\bibfnamefont{H.}~\bibnamefont{Happel}} \bibnamefont{and}
  \bibinfo{author}{\bibfnamefont{J.}~\bibnamefont{Brenner}},
  \emph{\bibinfo{title}{Low Reynolds number hydrodynamics with special
  pplications to particulate media}} (\bibinfo{publisher}{Martinus Nijhoff
  Publishers}, \bibinfo{address}{The Hague}, \bibinfo{year}{1983}).

\bibitem[{\citenamefont{Janoschek et~al.}(2013)\citenamefont{Janoschek,
  Harting, and Toschi}}]{Jens2013}
\bibinfo{author}{\bibfnamefont{F.}~\bibnamefont{Janoschek}},
  \bibinfo{author}{\bibfnamefont{J.}~\bibnamefont{Harting}}, \bibnamefont{and}
  \bibinfo{author}{\bibfnamefont{F.}~\bibnamefont{Toschi}},
  \bibinfo{journal}{arXiv preprint arXiv:1308.6482}  (\bibinfo{year}{2013}).

\bibitem[{\citenamefont{Amit}(2009)}]{Kumar_tesis}
\bibinfo{author}{\bibfnamefont{K.}~\bibnamefont{Amit}},
  \emph{\bibinfo{title}{TESIS}} (\bibinfo{publisher}{Chicago},
  \bibinfo{address}{Chicago}, \bibinfo{year}{2009}).

\bibitem[{\citenamefont{Cicuta et~al.}(2012)\citenamefont{Cicuta, Onofri,
  Lagomarsino, and Cicuta}}]{Cicuta2012}
\bibinfo{author}{\bibfnamefont{G.~M.} \bibnamefont{Cicuta}},
  \bibinfo{author}{\bibfnamefont{E.}~\bibnamefont{Onofri}},
  \bibinfo{author}{\bibfnamefont{M.~C.} \bibnamefont{Lagomarsino}},
  \bibnamefont{and} \bibinfo{author}{\bibfnamefont{P.}~\bibnamefont{Cicuta}},
  \bibinfo{journal}{Phys. Rev. E} \textbf{\bibinfo{volume}{85}},
  \bibinfo{pages}{016203} (\bibinfo{year}{2012}),
  \urlprefix\url{http://link.aps.org/doi/10.1103/PhysRevE.85.016203}.

\bibitem[{\citenamefont{Sierou and Brady}(2001)}]{BradySierou2001}
\bibinfo{author}{\bibfnamefont{A.}~\bibnamefont{Sierou}} \bibnamefont{and}
  \bibinfo{author}{\bibfnamefont{J.~F.} \bibnamefont{Brady}},
  \bibinfo{journal}{J. Fluid Mech.} \textbf{\bibinfo{volume}{448}},
  \bibinfo{pages}{115} (\bibinfo{year}{2001}).

\bibitem[{\citenamefont{Muldowney and Higdon}(1995)}]{MuldowneyHigdon1995}
\bibinfo{author}{\bibfnamefont{G.}~\bibnamefont{Muldowney}} \bibnamefont{and}
  \bibinfo{author}{\bibfnamefont{J.~J.~L.} \bibnamefont{Higdon}},
  \bibinfo{journal}{Journal of Fluid Mechanics} \textbf{\bibinfo{volume}{298}},
  \bibinfo{pages}{167–192} (\bibinfo{year}{1995}).

\bibitem[{\citenamefont{Dünweg and Ladd}(2009)}]{Duenweg2009}
\bibinfo{author}{\bibfnamefont{B.}~\bibnamefont{Dünweg}} \bibnamefont{and}
  \bibinfo{author}{\bibfnamefont{A.~J.~C.} \bibnamefont{Ladd}},
  \bibinfo{journal}{Advances in Polymer Science}
  \textbf{\bibinfo{volume}{221}}, \bibinfo{pages}{89} (\bibinfo{year}{2009}).

\bibitem[{\citenamefont{Chew et~al.}(2002)\citenamefont{Chew, Shu, and
  Peng}}]{Chew2002}
\bibinfo{author}{\bibfnamefont{Y.~T.} \bibnamefont{Chew}},
  \bibinfo{author}{\bibfnamefont{C.}~\bibnamefont{Shu}}, \bibnamefont{and}
  \bibinfo{author}{\bibfnamefont{Y.}~\bibnamefont{Peng}},
  \bibinfo{journal}{Journal of Statistical Physics}
  \textbf{\bibinfo{volume}{107}} (\bibinfo{year}{2002}).

\bibitem[{\citenamefont{Ferrás et~al.}(2013)\citenamefont{Ferrás, Nóbrega,
  and Pinho}}]{Ferras2013}
\bibinfo{author}{\bibfnamefont{L.}~\bibnamefont{Ferrás}},
  \bibinfo{author}{\bibfnamefont{J.}~\bibnamefont{Nóbrega}}, \bibnamefont{and}
  \bibinfo{author}{\bibfnamefont{F.}~\bibnamefont{Pinho}},
  \bibinfo{journal}{International Journal for Numerical Methods in Fluids}
  \textbf{\bibinfo{volume}{72}}, \bibinfo{pages}{724} (\bibinfo{year}{2013}),
  ISSN \bibinfo{issn}{1097-0363},
  \urlprefix\url{http://dx.doi.org/10.1002/fld.3765}.

\bibitem[{\citenamefont{Kapral}(2008)}]{Kapral_rev}
\bibinfo{author}{\bibfnamefont{R.}~\bibnamefont{Kapral}},
  \bibinfo{journal}{Advances in Chemical Physics}
  \textbf{\bibinfo{volume}{140}}, \bibinfo{pages}{89} (\bibinfo{year}{2008}).

\bibitem[{\citenamefont{Bian}(2014)}]{ElleroSplitting2014}
\bibinfo{author}{\bibfnamefont{M.}~\bibnamefont{Bian},
  \bibfnamefont{Xin;~Ellero}}, \bibinfo{journal}{Computer Physics
  Communications} \textbf{\bibinfo{volume}{185}}, \bibinfo{pages}{53}
  (\bibinfo{year}{2014}).

\bibitem[{\citenamefont{Nguyen and Ladd}(2002)}]{Nguyen2002}
\bibinfo{author}{\bibfnamefont{N.-Q.} \bibnamefont{Nguyen}} \bibnamefont{and}
  \bibinfo{author}{\bibfnamefont{A.}~\bibnamefont{Ladd}},
  \bibinfo{journal}{Physical Review E} \textbf{\bibinfo{volume}{66}},
  \bibinfo{pages}{046708} (\bibinfo{year}{2002}), ISSN
  \bibinfo{issn}{1063-651X},
  \urlprefix\url{http://link.aps.org/doi/10.1103/PhysRevE.66.046708}.

\bibitem[{\citenamefont{Uhlmann}(2005)}]{Uhlmann2005}
\bibinfo{author}{\bibfnamefont{M.}~\bibnamefont{Uhlmann}},
  \bibinfo{journal}{Journal of Computational Physics}
  \textbf{\bibinfo{volume}{209}}, \bibinfo{pages}{448} (\bibinfo{year}{2005}).

\bibitem[{\citenamefont{Breugem}(2010)}]{Breugem}
\bibinfo{author}{\bibfnamefont{W.-P.} \bibnamefont{Breugem}},
  \bibinfo{journal}{ASME 2010 3rd Joint US-European Fluids Engineering Summer
  Meeting and 8th International Conference on Nanochannels, Microchannels, and
  Minichannels FEDSM-ICNMM2010}  (\bibinfo{year}{2010}).

\bibitem[{\citenamefont{Feng and Michaelides}(2005)}]{Feng2005}
\bibinfo{author}{\bibfnamefont{Z.-G.} \bibnamefont{Feng}} \bibnamefont{and}
  \bibinfo{author}{\bibfnamefont{E.~E.} \bibnamefont{Michaelides}},
  \bibinfo{journal}{Journal of Computational Physics}
  \textbf{\bibinfo{volume}{202}}, \bibinfo{pages}{20} (\bibinfo{year}{2005}),
  ISSN \bibinfo{issn}{00219991},
  \urlprefix\url{http://linkinghub.elsevier.com/retrieve/pii/S0021999104002669%
}.

\bibitem[{\citenamefont{Molina and Yamamoto}(2013)}]{Molina-Yamamoto2013}
\bibinfo{author}{\bibfnamefont{J.}~\bibnamefont{Molina}} \bibnamefont{and}
  \bibinfo{author}{\bibfnamefont{R.}~\bibnamefont{Yamamoto}},
  \bibinfo{journal}{J Chem Phys.} \textbf{\bibinfo{volume}{139}},
  \bibinfo{pages}{234105} (\bibinfo{year}{2013}).

\bibitem[{\citenamefont{Peskin}(2002)}]{Peskin2002}
\bibinfo{author}{\bibfnamefont{C.}~\bibnamefont{Peskin}},
  \bibinfo{journal}{Acta Numerica} \textbf{\bibinfo{volume}{11}},
  \bibinfo{pages}{479} (\bibinfo{year}{2002}),
  \urlprefix\url{http://journals.cambridge.org/abstract\_S0962492902000077}.

\bibitem[{\citenamefont{Atzberger et~al.}(2007)\citenamefont{Atzberger, Kramer,
  and Peskin}}]{Atzberger2007}
\bibinfo{author}{\bibfnamefont{P.~J.} \bibnamefont{Atzberger}},
  \bibinfo{author}{\bibfnamefont{P.~R.} \bibnamefont{Kramer}},
  \bibnamefont{and} \bibinfo{author}{\bibfnamefont{C.~S.}
  \bibnamefont{Peskin}}, \bibinfo{journal}{Journal of Computational Physics}
  \textbf{\bibinfo{volume}{224}}, \bibinfo{pages}{1255} (\bibinfo{year}{2007}).

\bibitem[{\citenamefont{{Balboa Usabiaga} et~al.}(2014)\citenamefont{{Balboa
  Usabiaga}, Delgado-Buscalioni, Griffith, and Donev}}]{Balboa2014}
\bibinfo{author}{\bibfnamefont{F.}~\bibnamefont{{Balboa Usabiaga}}},
  \bibinfo{author}{\bibfnamefont{R.}~\bibnamefont{Delgado-Buscalioni}},
  \bibinfo{author}{\bibfnamefont{B.~E.} \bibnamefont{Griffith}},
  \bibnamefont{and} \bibinfo{author}{\bibfnamefont{A.}~\bibnamefont{Donev}},
  \bibinfo{journal}{Computer Methods in Applied Mechanics and Engineering}
  \textbf{\bibinfo{volume}{269}}, \bibinfo{pages}{139} (\bibinfo{year}{2014}).

\bibitem[{\citenamefont{{Balboa Usabiaga}}(2014)}]{Balboa_tesis}
\bibinfo{author}{\bibfnamefont{F.}~\bibnamefont{{Balboa Usabiaga}}},
  \emph{\bibinfo{title}{Minimal models for finite particles in fluctuating
  hydrodynamics}} (\bibinfo{publisher}{Universidad Autonoma de Madrid},
  \bibinfo{address}{Madrid}, \bibinfo{year}{2014}).

\bibitem[{\citenamefont{{Balboa Usabiaga} et~al.}(2012)\citenamefont{{Balboa
  Usabiaga}, Pagonabarraga, and Delgado-Buscalioni}}]{Balboa2012a}
\bibinfo{author}{\bibfnamefont{F.}~\bibnamefont{{Balboa Usabiaga}}},
  \bibinfo{author}{\bibfnamefont{I.}~\bibnamefont{Pagonabarraga}},
  \bibnamefont{and}
  \bibinfo{author}{\bibfnamefont{R.}~\bibnamefont{Delgado-Buscalioni}},
  \bibinfo{journal}{Journal of Computational Physics}
  \textbf{\bibinfo{volume}{235}}, \bibinfo{pages}{701} (\bibinfo{year}{2012}),
  ISSN \bibinfo{issn}{00219991},
  \urlprefix\url{http://linkinghub.elsevier.com/retrieve/pii/S0021999112006493%
}.

\bibitem[{\citenamefont{Bhalla et~al.}(2013)\citenamefont{Bhalla, Bale,
  Griffith, and Patankar}}]{Bhalla2013b}
\bibinfo{author}{\bibfnamefont{A.~P.~S.} \bibnamefont{Bhalla}},
  \bibinfo{author}{\bibfnamefont{R.}~\bibnamefont{Bale}},
  \bibinfo{author}{\bibfnamefont{B.~E.} \bibnamefont{Griffith}},
  \bibnamefont{and} \bibinfo{author}{\bibfnamefont{N.~A.}
  \bibnamefont{Patankar}}, \bibinfo{journal}{Journal of Computational Physics}
  \textbf{\bibinfo{volume}{250}}, \bibinfo{pages}{446 } (\bibinfo{year}{2013}),
  ISSN \bibinfo{issn}{0021-9991},
  \urlprefix\url{http://www.sciencedirect.com/science/article/pii/S00219991130%
03173}.

\bibitem[{\citenamefont{Felderhof}(2014)}]{Felderhof2014}
\bibinfo{author}{\bibfnamefont{B.~U.} \bibnamefont{Felderhof}},
  \bibinfo{journal}{Physical Review E} \textbf{\bibinfo{volume}{89}}
  (\bibinfo{year}{2014}).

\bibitem[{\citenamefont{{Balboa Usabiaga} and
  Delgado-Buscalioni}(2013)}]{Balboa_sound}
\bibinfo{author}{\bibfnamefont{F.}~\bibnamefont{{Balboa Usabiaga}}}
  \bibnamefont{and}
  \bibinfo{author}{\bibfnamefont{R.}~\bibnamefont{Delgado-Buscalioni}},
  \bibinfo{journal}{Physical Review E} \textbf{\bibinfo{volume}{88}}
  (\bibinfo{year}{2013}).

\bibitem[{\citenamefont{Roma et~al.}(1999)\citenamefont{Roma, Peskin, and
  Berger}}]{Roma1999}
\bibinfo{author}{\bibfnamefont{A.~M.} \bibnamefont{Roma}},
  \bibinfo{author}{\bibfnamefont{C.~S.} \bibnamefont{Peskin}},
  \bibnamefont{and} \bibinfo{author}{\bibfnamefont{M.~J.}
  \bibnamefont{Berger}}, \bibinfo{journal}{J. Comput. Phys.}
  \textbf{\bibinfo{volume}{153}}, \bibinfo{pages}{509} (\bibinfo{year}{1999}).

\bibitem[{\citenamefont{Zwanzig and Bixon}(1975)}]{Zwanzig1975}
\bibinfo{author}{\bibfnamefont{R.}~\bibnamefont{Zwanzig}} \bibnamefont{and}
  \bibinfo{author}{\bibfnamefont{M.}~\bibnamefont{Bixon}},
  \bibinfo{journal}{Journal of Fluid Mechanics} \textbf{\bibinfo{volume}{69}},
  \bibinfo{pages}{21} (\bibinfo{year}{1975}).

\bibitem[{\citenamefont{Landau and Lifshitz}(1959)}]{LandauFL}
\bibinfo{author}{\bibfnamefont{L.~D.} \bibnamefont{Landau}} \bibnamefont{and}
  \bibinfo{author}{\bibfnamefont{E.~M.} \bibnamefont{Lifshitz}},
  \emph{\bibinfo{title}{{Fluid Mechanics}}} (\bibinfo{publisher}{Pergamon
  Press, New York}, \bibinfo{year}{1959}).

\bibitem[{\citenamefont{Delong et~al.}(2014)\citenamefont{Delong, {Balboa
  Usabiaga}, Delgado-Buscalioni, Griffith, and Donev}}]{Stokeslimit}
\bibinfo{author}{\bibfnamefont{S.}~\bibnamefont{Delong}},
  \bibinfo{author}{\bibfnamefont{F.}~\bibnamefont{{Balboa Usabiaga}}},
  \bibinfo{author}{\bibfnamefont{R.}~\bibnamefont{Delgado-Buscalioni}},
  \bibinfo{author}{\bibfnamefont{B.~E.} \bibnamefont{Griffith}},
  \bibnamefont{and} \bibinfo{author}{\bibfnamefont{A.}~\bibnamefont{Donev}},
  \bibinfo{journal}{Journal of Chemical Physics}
  \textbf{\bibinfo{volume}{140}}, \bibinfo{pages}{134110}
  (\bibinfo{year}{2014}).

\bibitem[{not()}]{note}
\bibinfo{note}{In particular, $\mIc = \sum_{i,j} \bfs_i\cdot\bfs_j \left[
  \bs{1} \mathrm{Tr} - {\bs 1}\right] \bJ_i\bP\bS_j - \bJ_i\bP\bS_j :
  \bfs_i\bfs_j \bs{1} -\mathrm{Tr}(\bJ_i\bP\bS_j)\bfs_i\bfs_j
  +\bJ_i\bP\bS_j\cdot \bfs_i\bfs_j +\bfs_i\bfs_j\cdot\bJ_i\bP\bS_j$, where
  $\mathrm{Tr} {\bs A}=A^{\alpha,\alpha}$ is the trace operator}.

\bibitem[{\citenamefont{Pinelli et~al.}(2010)\citenamefont{Pinelli, , Naqavi,
  Piomelli, and Favier}}]{Pinelli2010a}
\bibinfo{author}{\bibfnamefont{A.}~\bibnamefont{Pinelli}}, ,
  \bibinfo{author}{\bibfnamefont{I.}~\bibnamefont{Naqavi}},
  \bibinfo{author}{\bibfnamefont{U.}~\bibnamefont{Piomelli}}, \bibnamefont{and}
  \bibinfo{author}{\bibfnamefont{J.}~\bibnamefont{Favier}},
  \bibinfo{journal}{Journal of Computational Physics}
  \textbf{\bibinfo{volume}{229}}, \bibinfo{pages}{9073} (\bibinfo{year}{2010}),
  ISSN \bibinfo{issn}{00219991},
  \urlprefix\url{http://linkinghub.elsevier.com/retrieve/pii/S0021999110004687%
}.

\bibitem[{\citenamefont{{Balboa Usabiaga} et~al.}(2013)\citenamefont{{Balboa
  Usabiaga}, Xie, Delgado-Buscalioni, and Donev}}]{BalboaStokesEinstein}
\bibinfo{author}{\bibfnamefont{F.}~\bibnamefont{{Balboa Usabiaga}}},
  \bibinfo{author}{\bibfnamefont{X.}~\bibnamefont{Xie}},
  \bibinfo{author}{\bibfnamefont{R.}~\bibnamefont{Delgado-Buscalioni}},
  \bibnamefont{and} \bibinfo{author}{\bibfnamefont{A.}~\bibnamefont{Donev}},
  \bibinfo{journal}{Journal of Chemical Physics}
  \textbf{\bibinfo{volume}{139}}, \bibinfo{pages}{214113}
  (\bibinfo{year}{2013}).

\bibitem[{\citenamefont{Maxey and Riley}(1983)}]{Maxey1983}
\bibinfo{author}{\bibfnamefont{M.~R.} \bibnamefont{Maxey}} \bibnamefont{and}
  \bibinfo{author}{\bibfnamefont{J.~J.} \bibnamefont{Riley}},
  \bibinfo{journal}{Physics of Fluids} \textbf{\bibinfo{volume}{26}},
  \bibinfo{pages}{883} (\bibinfo{year}{1983}).

\bibitem[{\citenamefont{Hasimoto}(1959)}]{Hasimoto}
\bibinfo{author}{\bibfnamefont{H.}~\bibnamefont{Hasimoto}},
  \bibinfo{journal}{J. Fluid Mech.} \textbf{\bibinfo{volume}{5}},
  \bibinfo{pages}{317} (\bibinfo{year}{1959}).

\bibitem[{\citenamefont{Mazur and Bedeaux}(1974)}]{Mazur1974}
\bibinfo{author}{\bibfnamefont{P.}~\bibnamefont{Mazur}} \bibnamefont{and}
  \bibinfo{author}{\bibfnamefont{D.}~\bibnamefont{Bedeaux}},
  \bibinfo{journal}{Physica} \textbf{\bibinfo{volume}{76}},
  \bibinfo{pages}{235} (\bibinfo{year}{1974}).

\bibitem[{\citenamefont{Zatovsky}(1969)}]{Zatovsky1969}
\bibinfo{author}{\bibfnamefont{A.~V.} \bibnamefont{Zatovsky}},
  \bibinfo{journal}{Izv. Vuzov. Fizika} \textbf{\bibinfo{volume}{10}}
  (\bibinfo{year}{1969}).

\bibitem[{\citenamefont{Rotne and Prager}(1969)}]{Rotne1969}
\bibinfo{author}{\bibfnamefont{J.}~\bibnamefont{Rotne}} \bibnamefont{and}
  \bibinfo{author}{\bibfnamefont{S.}~\bibnamefont{Prager}},
  \bibinfo{journal}{Journal of Chemical Physics} \textbf{\bibinfo{volume}{50}},
  \bibinfo{pages}{4831} (\bibinfo{year}{1969}).

\bibitem[{\citenamefont{Brenner}(1961)}]{Brenner1961}
\bibinfo{author}{\bibfnamefont{H.}~\bibnamefont{Brenner}},
  \bibinfo{journal}{Chem. Engng. Sci.} \textbf{\bibinfo{volume}{16}},
  \bibinfo{pages}{242} (\bibinfo{year}{1961}).

\bibitem[{\citenamefont{Hansford}(1970)}]{Hansford1970}
\bibinfo{author}{\bibfnamefont{R.~E.} \bibnamefont{Hansford}},
  \bibinfo{journal}{Matematika} \textbf{\bibinfo{volume}{17}},
  \bibinfo{pages}{270} (\bibinfo{year}{1970}).

\bibitem[{\citenamefont{Wagner and Brady}(2009)}]{PhysToday_WagnerBrady}
\bibinfo{author}{\bibfnamefont{N.~J.} \bibnamefont{Wagner}} \bibnamefont{and}
  \bibinfo{author}{\bibfnamefont{J.~F.} \bibnamefont{Brady}},
  \bibinfo{journal}{Physics Today} \textbf{\bibinfo{volume}{October}},
  \bibinfo{pages}{27} (\bibinfo{year}{2009}).

\bibitem[{\citenamefont{Hansen and McDonald}(1986)}]{Hansen_Simple_Liquids}
\bibinfo{author}{\bibfnamefont{J.~P.} \bibnamefont{Hansen}} \bibnamefont{and}
  \bibinfo{author}{\bibfnamefont{I.~R.} \bibnamefont{McDonald}},
  \emph{\bibinfo{title}{{Theory of Simple Liquids}}}
  (\bibinfo{publisher}{Academic Press}, \bibinfo{address}{New York},
  \bibinfo{year}{1986}).

\bibitem[{\citenamefont{Batchelor}(1967)}]{Batchelor}
\bibinfo{author}{\bibfnamefont{G.~K.} \bibnamefont{Batchelor}},
  \emph{\bibinfo{title}{An Introduction to Fluid Dynamics}}
  (\bibinfo{publisher}{Cambridge University Press},
  \bibinfo{address}{Cambridge}, \bibinfo{year}{1967}).

\bibitem[{\citenamefont{{Balboa Usabiaga}}()}]{Usabiaga}
\bibinfo{author}{\bibfnamefont{F.}~\bibnamefont{{Balboa Usabiaga}}},
  \emph{\bibinfo{title}{fluam, available at https://code.google.com/p/fluam/}}.

\bibitem[{\citenamefont{Yeo and Maxey}(2010)}]{Yeo2010}
\bibinfo{author}{\bibfnamefont{K.}~\bibnamefont{Yeo}} \bibnamefont{and}
  \bibinfo{author}{\bibfnamefont{M.~R.} \bibnamefont{Maxey}},
  \bibinfo{journal}{Journal of Computational Physics}
  \textbf{\bibinfo{volume}{229}}, \bibinfo{pages}{2401} (\bibinfo{year}{2010}).

\bibitem[{\citenamefont{de~Candia et~al.}(2009)\citenamefont{de~Candia, Gado,
  Fierro, and Coniglio}}]{Candia2009}
\bibinfo{author}{\bibfnamefont{A.}~\bibnamefont{de~Candia}},
  \bibinfo{author}{\bibfnamefont{E.~D.} \bibnamefont{Gado}},
  \bibinfo{author}{\bibfnamefont{A.}~\bibnamefont{Fierro}}, \bibnamefont{and}
  \bibinfo{author}{\bibfnamefont{A.}~\bibnamefont{Coniglio}},
  \bibinfo{journal}{Journal of Statistical Mechanics: Theory and Experiment} p.
  \bibinfo{pages}{P02052} (\bibinfo{year}{2009}).

\bibitem[{\citenamefont{Whitmer and Luijten}(2011)}]{Whitmer2011}
\bibinfo{author}{\bibfnamefont{J.~K.} \bibnamefont{Whitmer}} \bibnamefont{and}
  \bibinfo{author}{\bibfnamefont{E.}~\bibnamefont{Luijten}},
  \bibinfo{journal}{The Journal of Physical Chemistry B}
  \textbf{\bibinfo{volume}{115}}, \bibinfo{pages}{7294} (\bibinfo{year}{2011}).

\bibitem[{\citenamefont{Cao et~al.}(2012)\citenamefont{Cao, Cummins, and
  Morris}}]{Cao2012}
\bibinfo{author}{\bibfnamefont{X.}~\bibnamefont{Cao}},
  \bibinfo{author}{\bibfnamefont{H.}~\bibnamefont{Cummins}}, \bibnamefont{and}
  \bibinfo{author}{\bibfnamefont{J.}~\bibnamefont{Morris}},
  \bibinfo{journal}{Journal of colloid and interface science}
  \textbf{\bibinfo{volume}{368}}, \bibinfo{pages}{86} (\bibinfo{year}{2012}).

\end{thebibliography}

\end{document}